\documentclass[fleqn,usenatbib]{mnras}

\usepackage{newtxtext,newtxmath,bm,color}

\usepackage[T1]{fontenc}

\DeclareRobustCommand{\VAN}[3]{#2}
\let\VANthebibliography\thebibliography
\def\thebibliography{\DeclareRobustCommand{\VAN}[3]{##3}\VANthebibliography}

\usepackage{graphicx}	
\usepackage{amsmath}	

\title[Modeling stellar convective transport with plume]{Modeling stellar convective transport with plumes : II. Transport Properties of Locally and Non-locally driven Convection}

\author[Y. Masada et al.]{
Youhei Masada,$^{1}$\thanks{E-mail: ymasada@fukuoka-u.ac.jp}
Tomoya Takiwaki,$^{2}$
and Nobumitsu Yokoi$^{3}$
\\
$^{1}$Department of Applied Physics, Faculty of Science, Fukuoka University, Fukuoka 814-0180, Japan\\
$^{2}$Division of Science, National Astronomical Observatory of Japan (NAOJ), Osawa, Mitaka, Tokyo 181-8588, Japan\\
$^{3}$Institute of Industrial Science, University of Tokyo, Komaba, Meguro, Tokyo 153-8505, Japan
}

\date{Accepted 2026 October 14. Received 2026 September 15; in original form 2025 October 02}

\pubyear{2026}

\begin{document}
\label{firstpage}
\pagerange{\pageref{firstpage}--\pageref{lastpage}}
\maketitle

\begin{abstract}
We perform three-dimensional hydrodynamic simulations of two idealized convection setups: a cooling-driven model (Model~C) and an entropy-gradient–driven model (Model~S). The two setups exhibit clear differences: while Model~S develops large, relatively stationary eddies excited at depth, Model~C is dominated near the surface by intermittent plume-like downflows that produce broad non-Gaussian velocity distributions and a turbulent energy flux that exceeds that in Model~S by nearly an order of magnitude in the upper convection zone. Conventional gradient–diffusion (GD) closures reproduce the transport in Model~S but significantly underestimate it in Model~C, demonstrating that the plume-dominated transport realized in our cooling-driven setup cannot be reproduced by a local, gradient-based model. To address this, we introduce a Time–Space Double Averaging (TSDA) method that extracts coherent fluctuations, yielding a diagnostic variable $\tilde{\bm{u}}$ that peaks where the flux is largest. Building on this insight, we propose a modified GD closure in which the turbulent diffusivity is corrected by a plume-mediated term, achieving quantitative agreement with simulation results. Although the closure requires a calibrated model parameter and a careful choice of the averaging window, it provides a physically transparent framework that links coherent plume dynamics to mean-field transport and offers a pathway toward improved subgrid models for non-equilibrium stellar convection zones.
\end{abstract}



\section{Introduction}
The Sun, as our nearest star, serves as a natural laboratory for studying the extreme nature of plasma turbulence. Turbulence plays a crucial role in diverse solar processes, including coronal heating and large-scale activity phenomena such as the solar wind, solar flares, and coronal mass ejections (CMEs) \citep[e.g.,][]{cranmer+15}. Among these, the energy transport by convection-driven turbulence is particularly fundamental to the Sun’s dynamical evolution \citep[e.g.,][]{michaud+98,stix04}.

The energy released by nuclear fusion in the solar core is transported to the surface through two principal processes: radiative diffusion, which governs energy transport in the radiative zone, and convection-driven turbulence, which dominates in the outer layers known as the convection zone (CZ). In terms of these two principal contributions, the energy-flux balance at a radial position  \(r\) can be schematically written as
\begin{equation}
F_{\rm rad}+F_{\rm enth}\simeq \frac{L}{4\pi r^2} \;,
\end{equation}
where \(F_{\rm rad}\) and \(F_{\rm enth}\) denote the radiative and enthalpy fluxes, respectively, \(L\) is the luminosity, and \(r\) is the radial coordinate measured from the stellar center. Because the solar luminosity ($L_\odot$) is measured with high precision, this relation places a stringent constraint on the energy transport through the solar interior.

The enthalpy flux can in turn be characterized schematically by the correlation
\begin{equation}
F_{\rm enth}\propto \rho \langle \delta u_r \delta e_i\rangle \;,
\end{equation}
where $\rho$ is the density, $e_i$ is the specific internal energy, and the angular brackets denote an ensemble mean. 

For the ideal-gas equation of state adopted in this study, $P = (\gamma -1)\rho e_i$, the specific enthalpy is given by $h = e_i + P/\rho = \gamma e_i$.  Since $\gamma $ is constant in our simulations, the enthalpy fluctuation is directly proportional to the internal-energy fluctuation,
$\delta h = \gamma \delta e_i$. Accordingly, apart from the constant factor $\gamma$, the correlation $\langle \delta u_r \delta e_i\rangle $ provides a direct diagnostic of turbulent enthalpy transport.

This expression highlights that the second-order correlation $\langle \delta u_r \delta e_i\rangle$ is the key physical quantity for understanding turbulent transport in the solar CZ. This motivates a central question: \textit{what physical mechanisms determine the turbulent energy transport characterized by $\langle \delta u_r \delta e_i\rangle$ in the solar convection zone?}

The gradient–diffusion (GD) hypothesis, introduced by \citet{boussinesq77}, posits that the transport of scalar fluxes, such as the turbulent entropy flux $\delta {\bm u}\delta s$, occurs down the mean scalar gradient $\nabla s$, analogous to a diffusive process \citep[e.g.,][]{Pope00}. With this framework, the turbulent energy flux can be modeled as 
\begin{equation}
\langle \delta u_r \delta e_i\rangle = -\kappa_{T}\frac{\partial e_i}{\partial r} \;,
\end{equation}
where $\kappa_T$ is the turbulent transport coefficient (often called eddy diffusivity). Within the mixing-length (ML) hypothesis \citep[e.g.,][]{prandtl25,BV58,stix02}, the eddy diffusivity is further parameterized as
\begin{equation}
\kappa_{T} = C\sqrt{\delta u_r^2}\ l_{m} \;,
\end{equation}
where $C$ is a dimensionless constant and $l_{m}$ is the typical eddy size, or “mixing length.” It is worth noting that the expression of the turbulent energy flux, given by eqs. (3) and (4), can be derived from the fundamental equation without resorting to any heuristic modeling \citep[e.g.,][]{yokoi18}. Conventionally, the density scale height $H_\rho$ is adopted as $l_{m}$ in the solar/stellar CZ, since convective instability—driven locally by a negative entropy gradient (i.e., a super-adiabatic entropy distribution)—is characterized by $H_\rho$ \citep[see, e.g.,][]{hathaway84, spruit+90}. The GD–ML framework underpins modern stellar evolution theory and can be regarded as one of the foundational elements of astrophysics \citep[e.g.,][and references therein]{kippenhahn+90,joyce+23}.

This conventional GD–ML picture naturally portrays solar convection as a hierarchical, multi-scale phenomenon. In the standard solar model \citep[e.g.,][]{CD+96}, the CZ is fully super-adiabatic, with density varying by about six orders of magnitude across its depth \citep[e.g.,][]{stix04}. Because the local density scale height $H_\rho$ changes significantly with radius, the characteristic mixing length and hence the eddy size also vary across the CZ \citep[see, e.g.,][and references therein]{miesch05,miesch+08,rempel+11, rincon+18}.

At the solar surface, granules of about $1$ Mm in diameter and with lifetimes of $\sim$10 minutes represent the smallest observed injection scale. At intermediate scales, supergranules of $\sim$30 Mm and $\sim$20 hours lifetime are observed \citep[e.g.,][]{schrijver13}. Beyond these, “giant cells” are hypothesized to span the entire CZ, with scales of $\sim$200 Mm and lifetimes of about one month. Consequently, the convection spectrum is conventionally expected to range from small granules to expansive giant cells (see Fig. 1(b) for a schematic illustration).

However, recent observations have challenged this conventional picture. 
\citet{hathaway+15} showed that even in high-resolution, time-averaged spectra of surface flows, clear signatures of giant cells are difficult to detect. Similarly, \citet{greer+15} reported that the signatures of giant cells are weak and remain challenging to identify observationally \citep[see also][]{hathaway+21}. In addition, helioseismic measurements suggest that, especially at scales where giant cells are expected, the amplitude of convective velocities is more than an order of magnitude smaller than that predicted by global simulations of solar convection \citep[e.g.,][]{hanasoge+12, lord+14, proxauf+21, hanasoge22}. If large-scale convection is indeed as weak as these observations suggest, critical questions arise: how is solar luminosity transported, and how are the observed mean flows maintained? These unresolved issues constitute what is now widely known as the convection conundrum and represent a central challenge in modern solar interior physics \citep[see, e.g.,][]{hanasoge22, hotta+23, birch+24, stefan+25}.

In response to this convection conundrum, an alternative model has been discussed, originally suggested by \citet{rast+93,rieutord+95,spruit97} and \citet{rast98}, in which convection is primarily driven not by a pervasive super-adiabatic gradient but by surface radiative cooling. This scenario, which was later revisited in the context of the convection conundrum and numerically tested in detail by \citet[][hereafter CR16]{cossette+16}, treats the solar CZ as an open system: entropy is removed at the surface through radiation into space, creating dense, low-entropy plumes that drive vigorous downflows in the CZ. This cooling-driven model, initially proposed in the early 1990s, has recently received renewed attention as observational constraints have become more stringent \citep[see also,][]{brandenburg16,anders+17,nelson+18,anders+19,hanson+24,kapyla25}.

In summary, two contrasting paradigms for solar convection are currently under debate: the entropy-gradient–driven model and the cooling-driven model. In the former, the entire CZ is super-adiabatic, with mixing lengths tied to local scale heights, producing a multi-scale convection spectrum from granules to giant cells. In the latter, super-adiabaticity is restricted to near-surface layers, while the bulk of the CZ remains nearly adiabatic, with the maximum eddy size limited by the depth of the super-adiabatic zone.

The aim of this study is twofold: (1) to investigate, through three-dimensional numerical simulations, the distinct physical characteristics of cooling-driven and entropy-gradient–driven convection, and (2) to examine how turbulent energy transport is represented under the two contrasting driving configurations. While this work is motivated in part by the convection conundrum, our primary goal is not to resolve it directly, but rather to clarify the underlying physical mechanisms of convective transport in simplified settings. In particular, the imposed Newtonian cooling near the upper boundary is employed as a controlled means to generate strong, localized driving that leads to coherent plume formation, thereby enabling us to isolate and examine the role of nonlocal transport processes.

The remainder of this paper is organized as follows. \S~2 outlines the numerical setup, and \S~3 presents the simulation results contrasting cooling-driven and entropy-gradient–driven convection. In \S~4, building on our earlier study \citep{yokoi+22}, we develop a modified theoretical framework for turbulent energy transport that incorporates plume dynamics. \S~5 provides a discussion of the physical interpretation, limitations, and astrophysical implications of the results. Finally, \S~6 summarizes the main findings and highlights directions for future work.

\section{Numerical Setup}
\begin{figure*}
\includegraphics[width=0.95\textwidth]{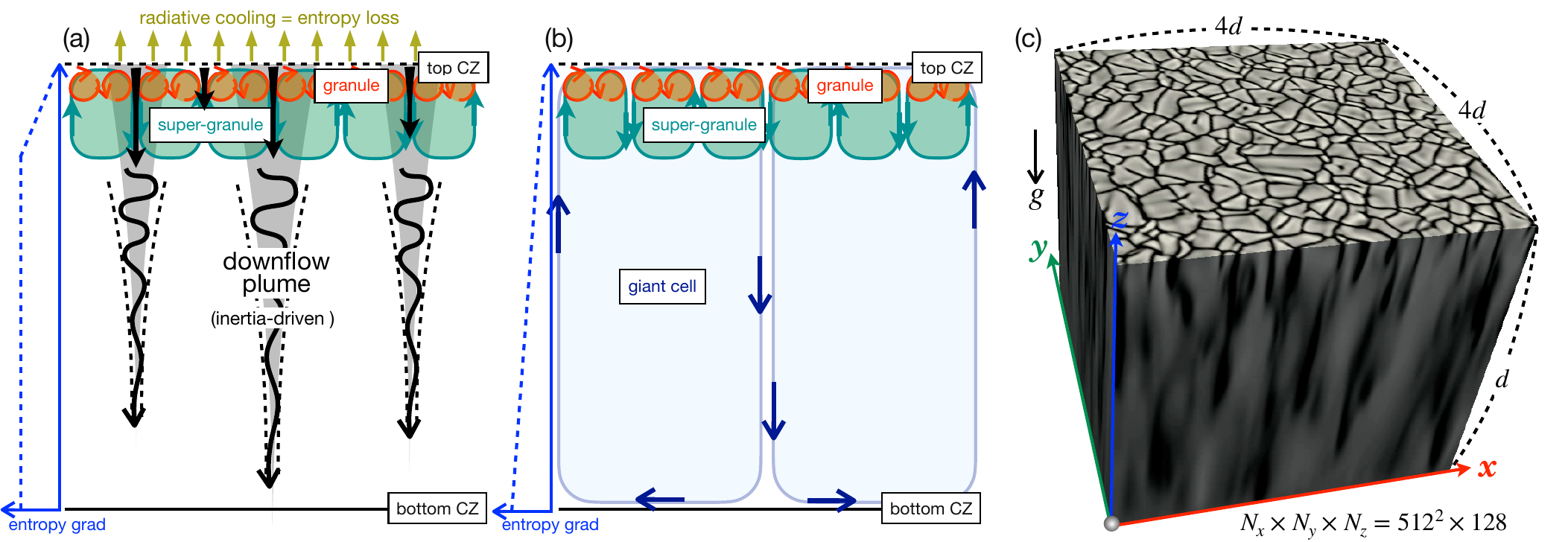}
\caption{Schematic illustrations of (a) the cooling-driven convection model (Model~C) and (b) the entropy-gradient–driven convection model (Model~S). Panel (c) shows a 3D visualization of the numerical setup adopted in this study and the typical convective state realized in our simulations (vertical velocity $u_z$ in Model~C is shown as a reference).}
\label{fig1}
\end{figure*}
\begin{figure}
\includegraphics[width=0.47\textwidth]{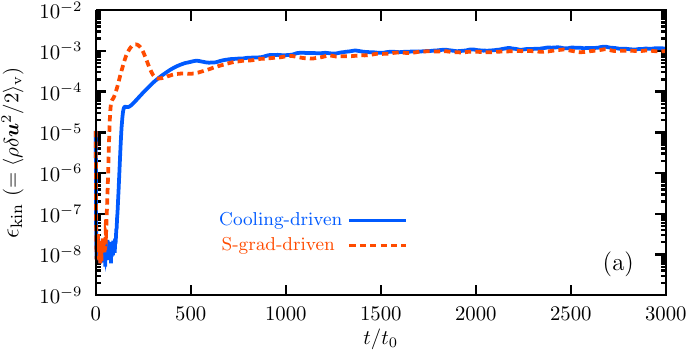}
\caption{Temporal evolution of volume-averaged kinetic energy, $\epsilon_{\rm kin} = \langle \rho \delta \bm{u}^2/2\rangle_{\rm v}$, for Model~C (blue solid) and Model~S (red dashed), where $\delta \bm{u} \equiv \bm{u} - \langle \bm{u} \rangle_{\rm h}$.}
\label{fig2}
\end{figure}
\begin{figure*}
\includegraphics[width=0.9\textwidth]{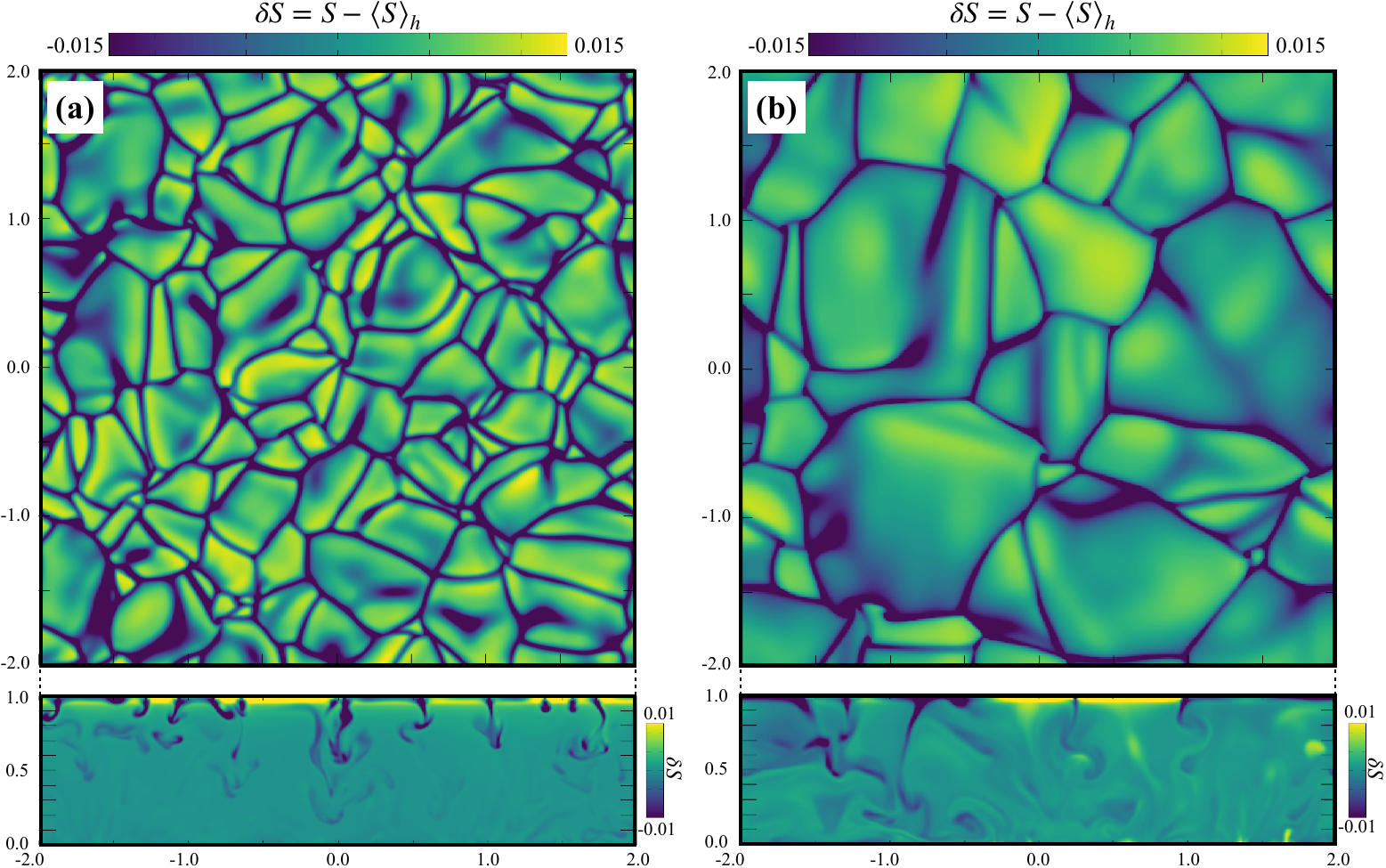}
\caption{Distribution of $\delta S$ on the horizontal cutting plane at $z = 0.95$ (top) and the vertical cutting plane at $y=0$ (bottom) for (a) Model~C and (b) Model~S, where $\delta S \equiv S - \langle S \rangle_{\rm h}$ is the fluctuating component of the entropy. The brighter (darker) tone corresponds to the region with a higher (lower) entropy than its horizontal average.}
\label{fig3}
\end{figure*}
We investigate turbulent energy transport in two convection models, illustrated in Fig.~1: (a) the cooling-driven model (Model~C) and (b) the entropy-gradient-driven (S-grad-driven) model (Model~S). To this end, we perform three-dimensional numerical simulations of the compressible hydrodynamic equations with explicit viscosity and thermal diffusion in a Cartesian coordinate system. Although the simulations are fully resolved numerically, the parameter regime is closer to that of large-eddy simulations (LES) than true DNS. The $x$- and $y$-axes are taken to be horizontal, while the $z$-axis is directed upward, anti-parallel to the gravity vector ${\bm g}$. The aspect ratio of the computational domain is $W/d=4$, common to both models, with box width $W$ ($x,y\in [-W/2,W/2]$) and depth $d$ ($z \in [0,d]$). 

The evolution of the system is determined by the equations of compressible hydrodynamics, which can be expressed in the form
\begin{eqnarray}
&&\frac{{\mathcal{D}} \rho}{\partial t}  =  - \rho \nabla\cdot {\bm u}  \;, \\ 
&&\frac{{\mathcal{D} }{\bm u}}{\mathcal{D} t}   =  - \frac{\nabla P}{\rho} + \frac{1}{\rho}\nabla \cdot (2\rho \nu_0 {\bm S}) + {\bm g}  \;, \\
&&\frac{\mathcal{D}e_i }{\mathcal{D} t}  =  - \frac{P\nabla\cdot {\bm u}}{\rho}  + \frac{\nabla\cdot(K_0 \nabla e_i)}{\rho} + 2\nu_0 \bm{S}^2 - \frac{e_i - e_{i,{\rm ref}}}{\tau}\;, 
\end{eqnarray}
with the strain rate tensor
\begin{equation}
  S_{ij} = \frac{1}{2}\left( \frac{\partial u_i}{\partial x_j} + \frac{\partial u_j}{\partial x_i}
  - \frac{2}{3}\delta_{ij}\frac{\partial u_k}{\partial x_k} \right) \;,   
\end{equation}
where $\mathcal{D}/\mathcal{D}t$ is the total derivative, $\rho$ is the density, $\bm u$ is the velocity, $P$ is the pressure, $e_i = c_{\rm V} T$ is the specific internal energy, and ${\bm g} = -g_0 {\bm e}_z$ represents a uniform gravity field with constant acceleration $g_0$. The viscosity and thermal conductivity are represented by $\nu_0$ and $K_0$, respectively. The other hydrodynamic variables have their usual meanings. Rotational effects, such as the Coriolis force, are ignored in this work. The last term in eq.~(7), the so-called Newtonian cooling, is imposed only in the surface region ($0.95 \le z/d \le 1$) of Model~C, where it relaxes $e_i$ toward the reference state $e_{i,{\rm ref}}$ and thereby mimics radiative cooling at the photosphere. We adopt a perfect gas law $P = (\gamma -1)\rho e_i$ with $\gamma = 5/3$ to close the system.

We assume an initial hydrostatic state with a polytropic stratification given by 
\begin{eqnarray}
&&e_i(z) = e_{i0} + \frac{g_0 (d-z)}{(\gamma-1)(m + 1)} \equiv e_{i,{\rm ref}}\;, \nonumber \\
&&\rho(z) = \rho_0 (e_i/e_{i0})^m \;,  \label{eq5}
\end{eqnarray}
where $m$ is the polytropic index, $e_{i0}$ and $\rho_0$ are the initial internal energy and density at the surface ($z=d$). 
The initial profile of $e_i$ is chosen as the reference state, i.e., $e_{i,{\rm ref}}$, for the Newtonian cooling in Model~C. 

The vertical profile of the polytropic index $m$ is slightly different between Models~C and S. In Model~C, $m=1.5$ is used for $0 \le z/d < 0.95$, except in the surface region ($0.95 \le z/d \le 1$) where a slightly super-adiabatic value $m=1.495$ is imposed. In Model~S, the super-adiabatic value $m=1.495$ is adopted throughout the whole domain. In both models, the degree of super-adiabaticity is identical, $\delta \equiv \nabla - \nabla_{\rm ad} = 8 \times 10^{-4}$, where $\nabla_{\rm ad} = 1 - 1/\gamma$ and $\nabla = 1/(m+1)$ for a polytropic atmosphere.

Since the condition $\delta > 0$ is satisfied in the whole domain in Model~S, the convection is driven by the local entropy ($S$) gradient. In contrast, in Model~C, the Newtonian cooling maintains the super-adiabaticity only in the surface region. Due to the loss of entropy through the cooling, fluid elements become heavier than the surrounding medium, leading to the convective motion dominated by downflow in the upper CZ. Once formed, the downflows gain inertia; even after penetrating the underlying adiabatic layer, they are not immediately braked by buoyancy. As a result, they descend in a plume-like manner, effectively transporting thermal energy from the deeper layers to the surface \citep[e.g.,][]{rast+93,spruit97, rast98, anders+17, nelson+18}. Our two models are similar to those studied by \citet[][CR16]{cossette+16}, except that CR16 included the stably stratified layer beneath the CZ, whereas we restrict the computational domain to the CZ itself. See Fig.~5 for the vertical profile of the superadiabaticity at the statistically steady state after convection has saturated.

Normalization is introduced by setting $g_0= \rho_0 = c_p = d = 1$. The normalized pressure scale-height at the surface,
defined by $\xi = H_p/d = (\gamma-1)e_{i0}/(g_0 d)$, controls the stratification level and is chosen here as $\xi = 0.019$,
yielding a relatively strong stratification with the density contrast between top and bottom CZs about $100$. 

All the physical variables are assumed to be periodic in horizontal directions. At the bottom and top boundaries, stress-free and impenetrable boundary conditions are imposed on the velocity, i.e., $\partial_z u_x = \partial_z u_y = u_z = 0$. A constant energy flux is imposed on the bottom 
boundary (i.e., $\partial_z e_i = {\rm const.}$), while the specific internal energy is fixed at the top boundary (i.e., $e_i = e_{i0}$). 

The basic equations are solved by the second-order Godunov-type finite-difference scheme that employs an approximate Riemann solver
\citep{sano+99,masada+14,bushby+18}. The simulations are performed with a Prandtl number ${\rm Pr}=1$, Rayleigh number ${\rm Ra}=4.2\times10^6$, and a grid resolution of $(N_x,N_y,N_z)=(512,512,128)$, where
the Prandtl and Rayleigh numbers are defined by 
\begin{equation}
  {\rm Pr}  =  \frac{\nu_0}{(K_0/\rho c_p )} \;, \ \ {\rm Ra}  =  \frac{g_0 d^3}{\chi_0 \nu_0}\frac{\delta}{\xi}  \;, \label{eq6}
\end{equation}
where $\rho$, $\delta$, and $\xi$ ($= H_p/d$) are evaluated at the surface ($z = d$). 

In the following, the volume- and horizontal-averages are denoted by angular brackets with subscripts ``\rm v", and ``\rm h", respectively. Time-averaging of these quantities is indicated by additional brackets. The convective turnover time is defined by $\tau_{\rm cv} \equiv 1/(u_{\rm cv}k_L)$, where $u_{\rm cv}$ is the volume-mean convective velocity at the saturated state and $k_L \equiv 2\pi/d$ represents the wavenumber associated with the largest energy-carrying eddies. See Fig.~1(c) for a 3D visualization of our numerical setup and the typical convective state obtained in our simulation ($u_z$ in Model~C is shown as a reference). 
\section{Numerical results}
\begin{figure}
\includegraphics[width=0.47\textwidth]{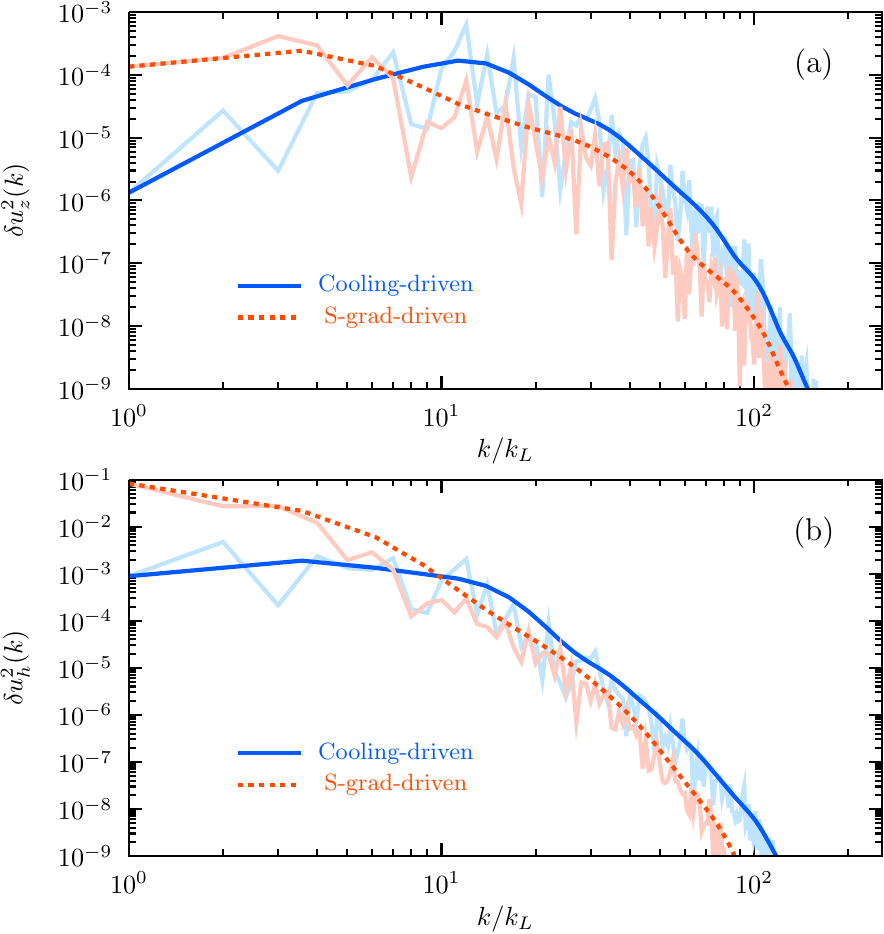}
\caption{Spectrum of (a) $\delta u_z^2$ and (b) $\delta u_h^2 (\equiv \delta u_x^2 + \delta u_y^2)$ for Model~C (blue solid) and Model~S (red dashed). The spectrum taken at each depth is projected onto a 1D wavenumber $k^2 = k_x^2 + k_y^2$ and then is averaged over the upper CZ, i.e., height from $z = 0.9$ to $0.95$. The pale colored line corresponds to the original spectrum, whereas the dark colored line represents the spectrum after the application of the Bezier smoothing.}
\label{fig4}
\end{figure}
\subsection{Basic properties of the convection}
In this subsection, we first characterize the temporal evolution and morphological features of convection in both models.

Figure~2 shows the temporal evolution of the volume-averaged kinetic energy, $\epsilon_{\rm kin} = \langle \rho \delta \bm{u}^2/2\rangle_{\rm v}$, for Model~C (blue solid) and Model~S (red dashed). Here $\delta \bm{u} \equiv \bm{u} - \langle \bm{u} \rangle_{\rm h}$ is the fluctuating component of the velocity, which practically coincides with $\bm{u}$ because the rotational effects are neglected in this study \citep[see, e.g.,][for mean-flow formation in rotating systems]{miesch+00}. Convection grows from the initial perturbations and saturates after $t \sim 80\tau_{\rm cv}$, where $\tau_{\rm cv}=16$ in Model~C and $\tau_{\rm cv}=19$ in Model~S. Although the growth histories differ somewhat, the saturated levels of kinetic energy are nearly the same in both cases. In the following, we focus on the quasi-equilibrium states. 

Figure~3 displays entropy fluctuations $\delta S \equiv S - \langle S \rangle_{\rm h}$ on horizontal cutting planes at $z=0.95$ (top panels) and vertical cutting planes at $y=0$ (bottom panels) for (a) Model~C and (b) Model~S. Bright (dark) regions indicate higher (lower) entropy relative to the horizontal mean. Both models exhibit the well-known up-down asymmetry of compressible convection: broader, slower upflows with positive $\delta S$, and narrower, faster downflows with negative $\delta S$ \citep[e.g.,][and references therein]{hurlburt+84, stein+98, miesch+00, brummell+02, nordlund+09, miesch+09}.

Despite these common features, the overall morphology differs markedly. In Model~C, plume-like downflows are spontaneously generated near the surface, whereas large-scale convective flows from deeper layers are absent, leading to a predominance of small-scale convective cells. In contrast, in Model~S, convection is driven throughout the domain by the entropy gradient, exciting large-scale convective flows at depth whose influence is imprinted even near the surface.

To quantify the difference in convective scales, we compare spectra of vertical velocity. Figure~4 shows Fourier spectra of (a) $\delta u_z^2$ and (b) $\delta u_h^2 (\equiv \delta u_x^2 + \delta u_y^2)$ for Model~C (blue solid) and Model~S (red dashed). Here the spectrum taken at each depth is projected onto a 1D wavenumber $k^2 = k_x^2 + k_y^2$ and then is averaged over the upper CZ, i.e., height from $z = 0.9$ to $0.95$. The pale curves denote raw spectra, while the dark curves are smoothed. Focusing on $\delta u_z^2$, Model~C peaks at $k/k_L\simeq 15$, whereas Model~S is dominated by modes with $k/k_L \lesssim 4$, consistent with the 2D horizontal patterns of the convective velocity in Fig.~3.

The typical scale of convective cells can be related to the local pressure scale height $H_p=(\gamma-1)e_i/g_0$, where $e_i$ has a height dependence (see eq.~(\ref{eq5})) \citep[e.g.,][]{chandra+61,spiegel+71, hathaway+83}. Therefore, from the peak wavenumber in the spectrum, we can extract information about the depth at which the energy-carrying vortices are being driven. Defining $k_p \equiv 2\pi/H_p$, we can evaluate $k_p/k_L \simeq 17$ at $z=0.9$ (upper CZ), or $k_p/k_L \simeq 3$ at $z=0.1$ (lower CZ), implying that Model~C is governed by vortices generated in the upper CZ, while in Model~S the energy transport is dominated by larger eddies originating in the lower CZ. These results are consistent with the earlier findings of CR16. 
\subsection{Thermodynamic structure and energy transport}
\begin{figure}
\includegraphics[width=0.47\textwidth]{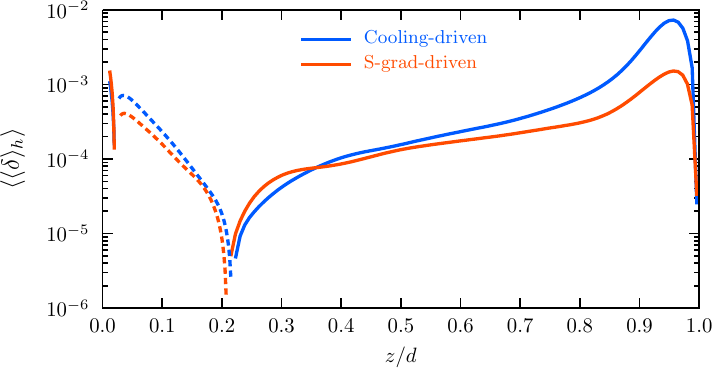}
\caption{Vertical profile of the time- and horizontally averaged superadiabaticity, $\langle\langle \delta \rangle_{\rm h}\rangle$ ($\delta \equiv \nabla - \nabla_{\rm ad}$), in the statistically steady state for Model~C (blue) and Model~S (red). The profiles are shown on a logarithmic scale over the full CZ in order to highlight differences in weakly superadiabatic regions. Solid lines denote positive values ($\delta>0$), while dashed lines denote the absolute value of negative superadiabaticity ($|\delta|$ for $\delta<0$). }
\label{fig5}
\end{figure}

\begin{figure}
\includegraphics[width=0.47\textwidth]{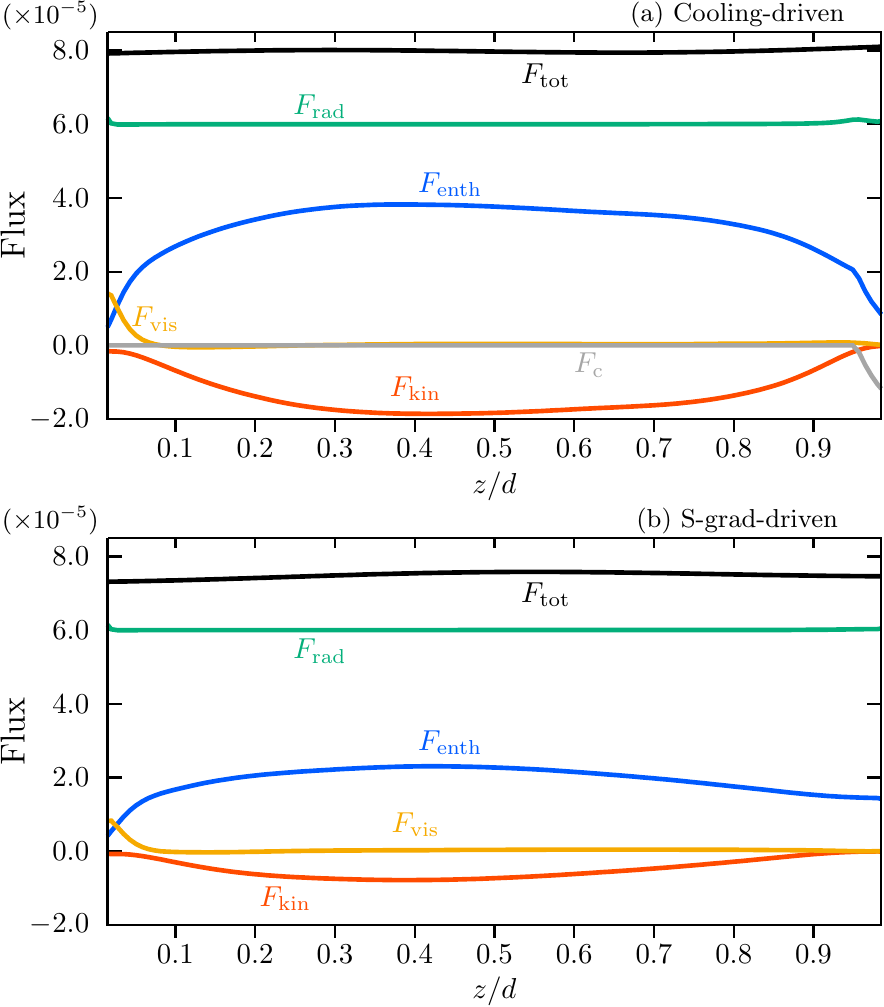}
\caption{Time- and horizontally averaged vertical energy fluxes in the statistically steady state for (a) Model C and (b) Model S. \(F_{\rm enth}\) is the enthalpy flux (blue), \(F_{\rm rad}\) is the radiative (diffusive) flux (green), \(F_{\rm kin}\) is the kinetic-energy flux (red), \(F_{\rm vis}\) is the viscous transport flux defined by eq.~(16) (orange), \(F_{\rm c}\) is the cumulative contribution associated with Newtonian cooling (gray; Model C only), and \(F_{\rm tot}\) is the total diagnostic flux (black).}
\label{fig6}
\end{figure}

To clarify how the two driving configurations differ after nonlinear saturation, we examine the thermodynamic structure and energy transport in the statistically steady state.

Figure 5 shows the vertical profile of the time- and horizontally averaged superadiabaticity, $\langle\langle \delta \rangle_{\rm h}\rangle$, where $\delta \equiv \nabla - \nabla_{\rm ad}$, in the statistically steady state for Model~C (blue) and Model~S (red). 
Here, positive values of $\delta$ are shown by solid lines, while negative values are plotted as dashed lines using their absolute values.

The equilibrated profiles reveal systematic differences between the two models. Model C exhibits a stronger near-surface enhancement of superadiabaticity associated with the imposed cooling layer, whereas Model S maintains a weakly superadiabatic structure over a broader depth range. Thus, the main difference is not a sharp separation of the superadiabatic layer, but the relative concentration of the driving toward the surface in Model C compared with the more distributed driving in Model S.

To highlight the relative contributions of each transport process of energy, Figure 6 presents the vertical profiles of time- and horizontally averaged energy flux \citep[e.g.,][]{brun+04,miesch+08}. For the analysis of the vertical energy transport, we define the total diagnostic flux, $F_{\rm tot}$,  as
\begin{equation}
F_{\rm tot} = F_{\rm enth} + F_{\rm kin} + F_{\rm rad} + F_{\rm vis} + F_{\rm c}\;,
\end{equation}
where the first four terms represent resolved transport fluxes, while $F_{\rm c}$ is a cumulative contribution associated with the imposed cooling term, introduced so that the stationary energy balance can be expressed in flux form. 

The turbulent enthalpy flux is defined as
\begin{equation}
F_{\rm enth} = \left\langle \left\langle \rho u_z \left( h-\langle h\rangle_{\rm h} \right) \right\rangle_{\rm h} \right\rangle ,
\end{equation}
where
\begin{equation}
h=e_i+\frac{p}{\rho}
\end{equation}
is the specific enthalpy. This definition isolates the turbulent enthalpy transport associated with correlations between vertical
motions and enthalpy fluctuations. In the limit of vanishing mean vertical mass flux, it is equivalent to the fluctuating part of the full enthalpy flux.

The kinetic-energy flux is
\begin{equation}
F_{\rm kin} = \left\langle \left\langle \frac{1}{2}\rho {\bm u}^{2}u_z \right\rangle_{\rm h} \right\rangle .
\end{equation}

The radiative (diffusive) flux is evaluated consistently with the thermal-diffusion term in the internal-energy equation,
\begin{equation}
F_{\rm rad} = - \left\langle \left\langle K_0\frac{\partial e_i}{\partial z} \right\rangle_{\rm h} \right\rangle ,
\end{equation}
where \(K_0\) is the thermal conductivity and \(e_i\) is the specific internal energy.

The viscous contribution to the transport flux is the mechanical transport associated with the viscous stress,
\begin{equation}
F_{\rm vis} = - \left\langle \left\langle u_i\tau_{iz} \right\rangle_{\rm h} \right\rangle ,
\end{equation}
where
\begin{equation}
\tau_{ij}=2\rho\nu_0 S_{ij}
\end{equation}
is the viscous stress tensor, with \(S_{ij}\) defined in eq.~(8). Although viscous dissipation,
\begin{equation}
Q_{\rm visc}=2\rho\nu_0 S_{ij}S_{ij},
\end{equation}
appears as a source term in the internal-energy equation, it represents conversion of kinetic energy into internal energy and therefore does not constitute an additional flux in the total-energy budget. Accordingly, no vertically integrated viscous-heating term is included in \(F_{\rm vis}\).

The Newtonian cooling term is a local source or sink of energy rather than a genuine transport flux. For convenience in examining the vertically integrated energy balance, we express its cumulative contribution in flux form as
\begin{equation}
F_{\rm c}(z) = - \int_{z_{\rm bot}}^{z} \left\langle Q_{\rm cool}(z')\right\rangle_{\rm h,t}\,dz',
\end{equation}
where
\begin{equation}
Q_{\rm cool} =
-\rho\frac{e_i-e_{i,\rm ref}}{\tau}.
\end{equation}
With this definition,
\begin{equation}
\frac{dF_{\rm c}}{dz} = -\langle Q_{\rm cool}\rangle_{\rm h,t},
\end{equation}
and hence $F_{\rm tot}$ is expected to be independent of height in a statistically stationary state, provided that all contributions to the energy transport are fully captured.

The total flux is nearly constant with depth in both models, indicating that the vertical energy transport is well balanced. The temporal evolution of the mean thermal state is shown in Appendix A, confirming that the statistical analysis is performed after both simulations have reached quasi-steady thermal states.

Despite the similar total energy transport, the partitioning among the flux components differs substantially between the two models. In both cases, the radiative flux provides the dominant contribution throughout the layer, forming a baseline energy transport. 

On top of this background, the enthalpy flux exhibits markedly different behavior. In Model~C, the enthalpy flux is significantly enhanced across the convection zone, providing a substantial additional contribution to the total transport. In contrast, Model~S shows a much weaker enthalpy flux, so that the energy transport remains largely diffusive.

This difference is particularly noteworthy because it is not directly reflected in the velocity power spectrum. Although the large-scale velocity amplitude is not necessarily larger in the cooling-driven case, the enthalpy flux is significantly enhanced. This reflects the fact that the enthalpy flux depends on the correlation between vertical velocity and thermodynamic fluctuations, rather than on the velocity amplitude alone. The cooling-driven setup produces strongly asymmetric convective structures associated with localized surface cooling, which leads to more efficient convective heat transport.

These results demonstrate that the two setups do not simply relax to identical stratifications, but retain distinct thermodynamic and transport characteristics associated with their different driving mechanisms. We therefore regard Models C and S as two contrasting convection setups, rather than as evidence for a sharp regime transition.
\subsection{Statistical properties of the convection: probability density}
\begin{figure}
\includegraphics[width=0.47\textwidth]{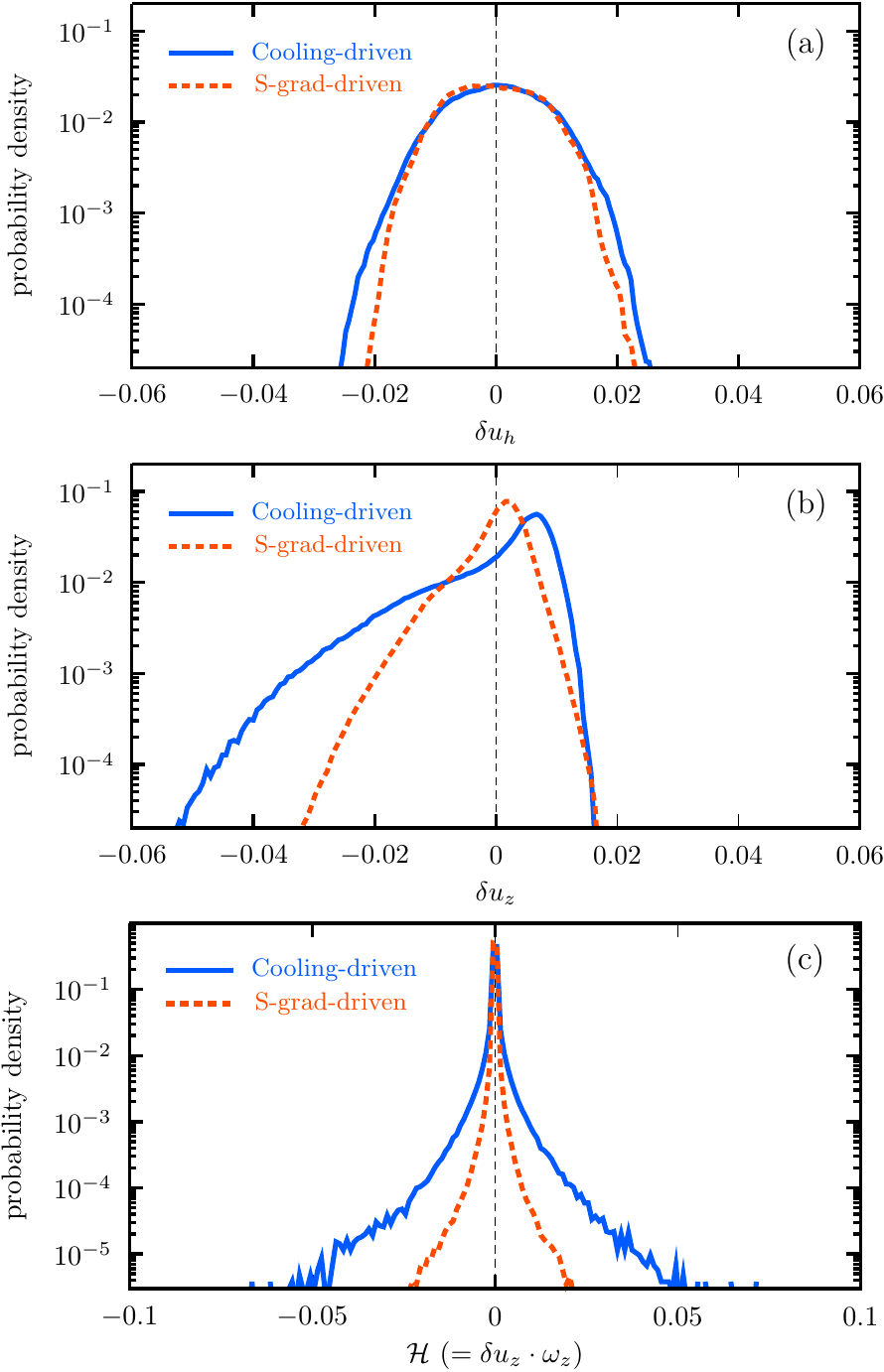}
\caption{Probability density functions on a horizontal plane at $z=0.95$ for (a) horizontal velocity $\delta u_h$, (b) vertical velocity $\delta u_z$, and (c) kinetic helicity $\mathcal{H} \equiv \delta u_z \omega_z$. The PDFs are normalized such that their integrals equal unity; the PDF therefore has units inverse to those of the corresponding variable. Results are shown for Model~C (blue solid) and Model~S (red dashed). While $\delta u_h$ exhibits similar Gaussian-like distributions in both models, clear differences appear in $\delta u_z$, reflecting broader wings in Model~C due to intermittent plume-like downflows.}
\label{fig7}
\end{figure}
\begin{figure}
\includegraphics[width=0.47\textwidth]{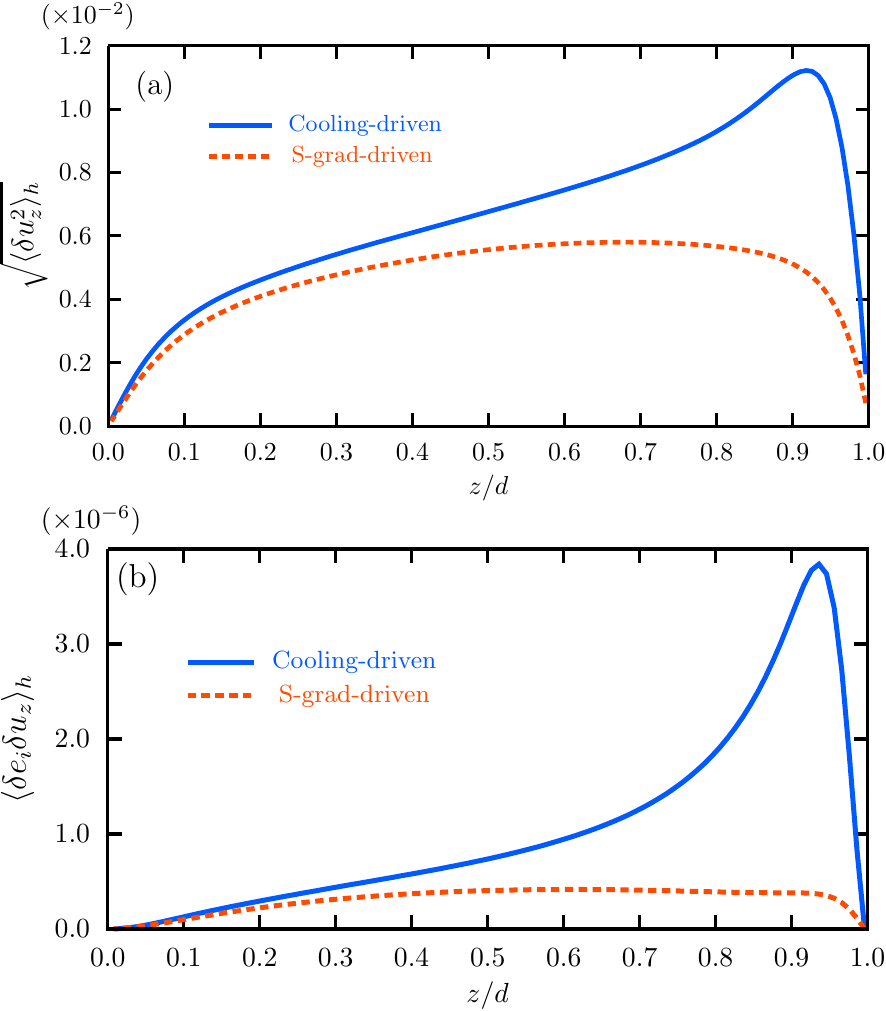}
\caption{Vertical profiles of (a) $\sqrt{\langle \delta u_z^2\rangle_{\rm h}}$, the RMS turbulent velocity in the vertical direction, and (b) $\langle \delta e_i \delta u_z \rangle_{\rm h}$, the mean turbulent energy flux, for Model~C (blue solid) and Model~S (red dashed). Model~C exhibits stronger vertical velocities and significantly enhanced energy flux in the upper convection zone compared to Model~S.}
\label{fig8}
\end{figure}
\begin{figure*}
\includegraphics[width=0.95\textwidth]{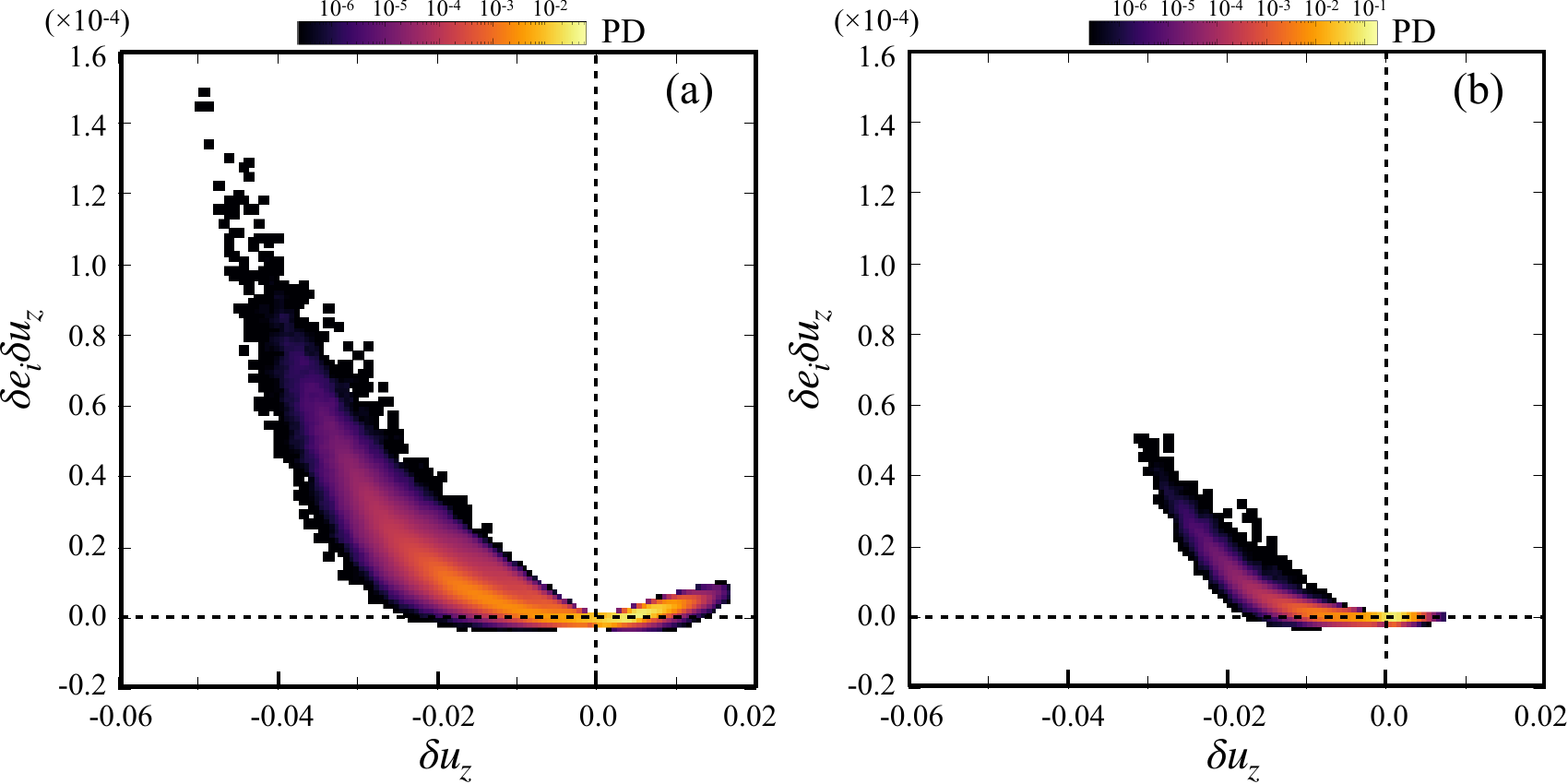}
\caption{Joint probability density distributions of the vertical velocity ($\delta u_z$) and vertical turbulent energy flux ($\delta e_i \delta u_z$) at the CZ surface ($z = 0.95$)} for (a) Model~C and (b) Model~S. In both cases, downflows dominate the transport, but Model~C exhibits a broader distribution associated with intermittent plume-like events.
\label{fig9}
\end{figure*}

\begin{figure}
\includegraphics[width=0.4\textwidth]{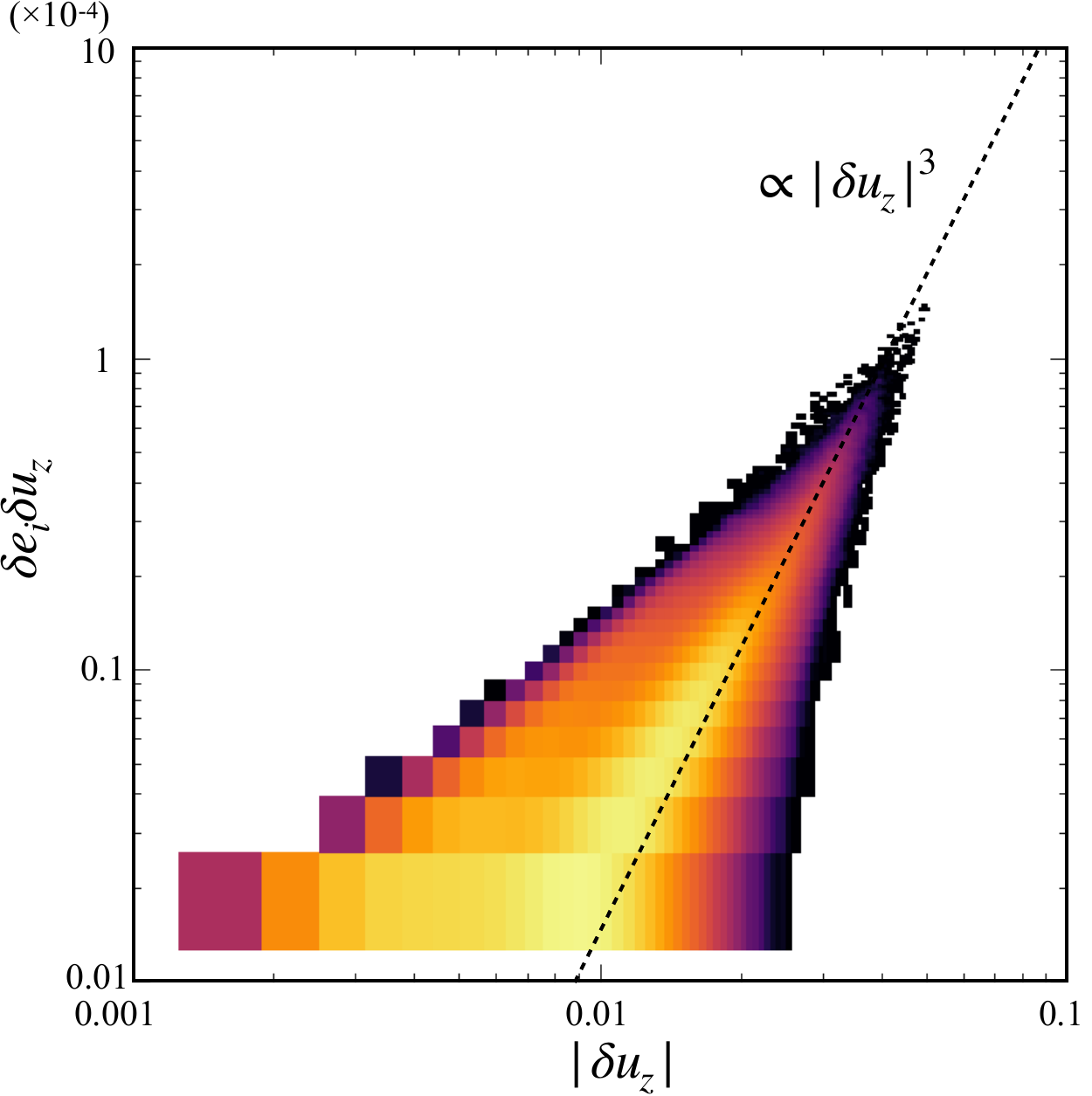}
\caption{Same as Fig.~9(a), but shown with both axes in logarithmic scale, focusing on the downflow regime ($\delta u_z < 0$). The dashed line indicates the reference scaling $\delta e_i \delta u_z \propto |\delta u_z|^3$, highlighting the nonlinear relation between velocity and energy transport in plume-dominated convection.}
\label{fig10}
\end{figure}

We next investigate how the statistical properties of convection differ between the two models. Figure 7 shows probability density functions evaluated on the horizontal cutting plane at \(z=0.95\) for (a) horizontal velocity $\delta u_h$, (b) vertical velocity $\delta u_z$, and (c) kinetic helicity defined by $\mathcal{H} \equiv \delta u_z \cdot \omega_z$, where $\omega_z = \partial_x u_y - \partial_y u_x$ is the vertical component of vorticity. The blue solid curve corresponds to Model~C and the red dashed curve to Model~S in each panel. 

For $\delta u_h$, the distribution is nearly Gaussian in both models, with little difference between them (Fig.~7a). In contrast, the probability density of $\delta u_z$ reveals a clear model dependence. Both cases exhibit the up–down asymmetry characteristic of compressible convection and mass-flux conservation \citep[e.g.,][]{hurlburt+84, stein+98, brummell+02}, but the downflow wing ($\delta u_z < 0$) is substantially broader in Model~C than in Model~S (Fig.~7b). This reflects the stochastic generation of plume-like downflows driven by surface radiative cooling in Model~C \citep[see, e.g.,][for the probability density of the compressible convection]{ahlers+09,he+19,kumar+22}.

Even in a non-rotating system, individual convective vortices can carry kinetic helicity locally (Fig.~7c). However, the ensemble-averaged helicity vanishes in both models, implying that no large-scale flow is induced. Nevertheless, the probability distributions demonstrate that convection in Model~C, in particular, is highly non-Gaussian due to the intermittent plume-like downflows. See also Figs.~9 and ~10, which further illustrate the non-Gaussianity arising from plume-driven downflows. These results indicate that the mean energy transport cannot be characterized by simple Gaussian statistics, thereby motivating the development of a theoretical framework, which we present in \S4.
\subsection{Mean properties of the convection: turbulent energy flux}
As shown in \S~3.1 and \S~3.2, the basic properties of convection differ substantially between the two models. Here we examine how these differences affect the vertical turbulent energy transport.

Hereafter, we use \(\langle\delta e_i\delta u_z\rangle_h\) as a diagnostic of turbulent energy transport. This quantity is closely related to, but distinct from, the density-weighted enthalpy flux \(F_{\rm enth}\) introduced in \S~3.2.

Figure~8 shows the vertical profiles of (a) the RMS vertical velocity, $\sqrt{\langle \delta u_z^2 \rangle_{\rm h}}$, and (b) the mean turbulent energy flux, $\langle \delta e_i \delta u_z \rangle_{\rm h}$, for Model~C (blue solid) and Model~S (red dashed). In the lower part of the CZ, the RMS velocity is nearly identical between models, indicating that the deep dynamics are comparable (Fig.~8a). However, in the upper CZ, the vertical velocity in Model~C exceeds that in Model~S by nearly a factor of two, reflecting the contribution of plume-like downflows that are stochastically generated by surface cooling. Although the strongest differences appear near the surface, Fig.~8 shows that the discrepancy between Model~C and Model~S persists well into the mid-CZ. This suggests that the discrepancy cannot be explained solely by a local boundary effect and is consistent with nonlocal plume-driven transport.

The difference is even more pronounced in the turbulent energy flux (Fig.~8b). In the upper CZ, the flux in Model~C is almost an order of magnitude larger than in Model~S, implying that intermittent plumes play a dominant role in transporting internal energy upward. In Model~S, by contrast, the transport is mediated by broader, large-scale circulations that originate from deeper layers, leading to smoother and less intense fluxes in the upper CZ. 

The nature of these transports can be further clarified by joint probability distributions. Figure~9 shows 2D histograms of vertical velocity $\delta u_z$ and turbulent energy flux $\delta e_i \delta u_z$ for (a) Model~C and (b) Model~S at the CZ surface ($z = 0.95$). In both models, downflows ($\delta u_z<0$) dominate the net flux. Yet, in Model~C, the probability density is spread over a much wider range, and faster downflows are clearly associated with stronger transport. This intermittent behavior reflects the non-equilibrium character of plume-driven convection.

Figure~10, which focuses on the downflow regime of Model~C in logarithmic scale, reveals a striking scaling relation: the turbulent energy flux is approximately proportional to $\delta u_z^3$. This cubic dependence suggests that the flux is controlled by strongly nonlinear processes associated with plumes, rather than by a diffusive process proportional to local gradients.

These results highlight a marked difference between the two driving configurations: in Model~C, turbulent energy transport is dominated by stochastic, plume-driven downflows, yielding intermittent and highly non-Gaussian flux statistics (see also Figs.~9 and 10). In Model~S, the transport is governed by larger-scale eddies excited in the lower CZ, resulting in smoother and more Gaussian-like behavior. While CR16 pointed out qualitative differences between cooling-driven and gradient-driven convection, the pronounced non-Gaussianity of plume-driven flows and the associated enhancement of turbulent energy flux have not been investigated before. These aspects represent new findings of the present study.

The scaling relation $\langle \delta e_i \delta u_z \rangle \propto \delta u_z^3$ observed in Model~C suggests that plume-driven transport may not be adequately captured by simple gradient-diffusion closures such as mixing-length theory. This motivates the development of an extended theoretical framework for turbulent energy transport, which we present in \S4. 

Further insight into the origin of this asymmetry is provided in Appendix B, where we examine additional diagnostics such as the skewness of the vertical velocity and the separate contributions of upflows and downflows to the turbulent transport.
\section{Modeling of Convective Energy Transports}
We now consider a closure model for the turbulent energy flux \(\langle\delta e_i\delta u_z\rangle_{\rm h}\) introduced in \S~3.4.
\subsection{Comparison with Gradient Diffusion Models}
A common approach to modeling turbulent transport in stellar CZs is the gradient-diffusion (GD) approximation, in which turbulent fluxes are modeled as proportional to the gradient of the corresponding mean quantity:
\begin{equation}
\langle \delta e_i \delta u_z \rangle_{\rm h} \simeq -\kappa_{\rm E} \frac{\partial \langle e_i \rangle_{\rm h}}{\partial z},
\end{equation}
where $\kappa_{\rm E}$ is the turbulent transport coefficient, often parameterized in terms of a characteristic velocity scale ($u$) and correlation length ($l$), i.e., $\kappa_{\rm E} = C u l$, where $C$, often referred to as the mixing-length parameter, is a dimensionless constant of arbitrary magnitude \citep[e.g.,][]{bradshaw74,boussinesq77,fiedler88,Pope00}. 
Following the conventional mixing-length approach, we adopt the RMS velocity, $u_{\rm rms}$, as the characteristic velocity scale $u$ and the vertical extent of the box, $d$, as the typical length scale $l$, although choosing the local density scale height $H_\rho$ instead does not affect the main conclusions.

Figure~11 compares the vertical profiles of the turbulent energy flux measured in our simulations (blue/red curves) with the predictions from the GD model (black dashed curves), obtained directly from the simulated mean profiles via eq.~(22). Panel~(a) corresponds to the cooling-driven convection (Model~C) and panel~(b) to the entropy-gradient-driven convection (Model~S). For Model~S, the GD model reproduces the simulated flux reasonably well across most of the CZ. This indicates that, when convection is excited by a pervasive entropy gradient, the transport process is well captured by a local closure that links the flux to the mean gradient.

By contrast, Model~C shows a pronounced deviation from GD predictions. In the upper CZ, where radiative cooling induces intermittent, plume-like downflows, the simulated flux is almost an order of magnitude larger than the GD prediction. The GD model underestimates transport in this region because the mean entropy gradient is small, even though vigorous plumes carry substantial energy. This discrepancy clearly demonstrates that the cooling-driven convection cannot be described by a simple gradient-diffusion closure. Instead, the flux is controlled by nonlocal, plume-dominated processes that are not reducible to the mean gradient.

In summary, the GD model is suitable for describing energy transport in the entropy-gradient-driven setup (Model~S), but fails in the cooling-driven setup (Model~C). These results highlight the need for an alternative modeling framework that explicitly accounts for the role of intermittent plumes in mediating turbulent transport. We will address this issue in the following subsections.
\begin{figure}
\includegraphics[width=0.47\textwidth]{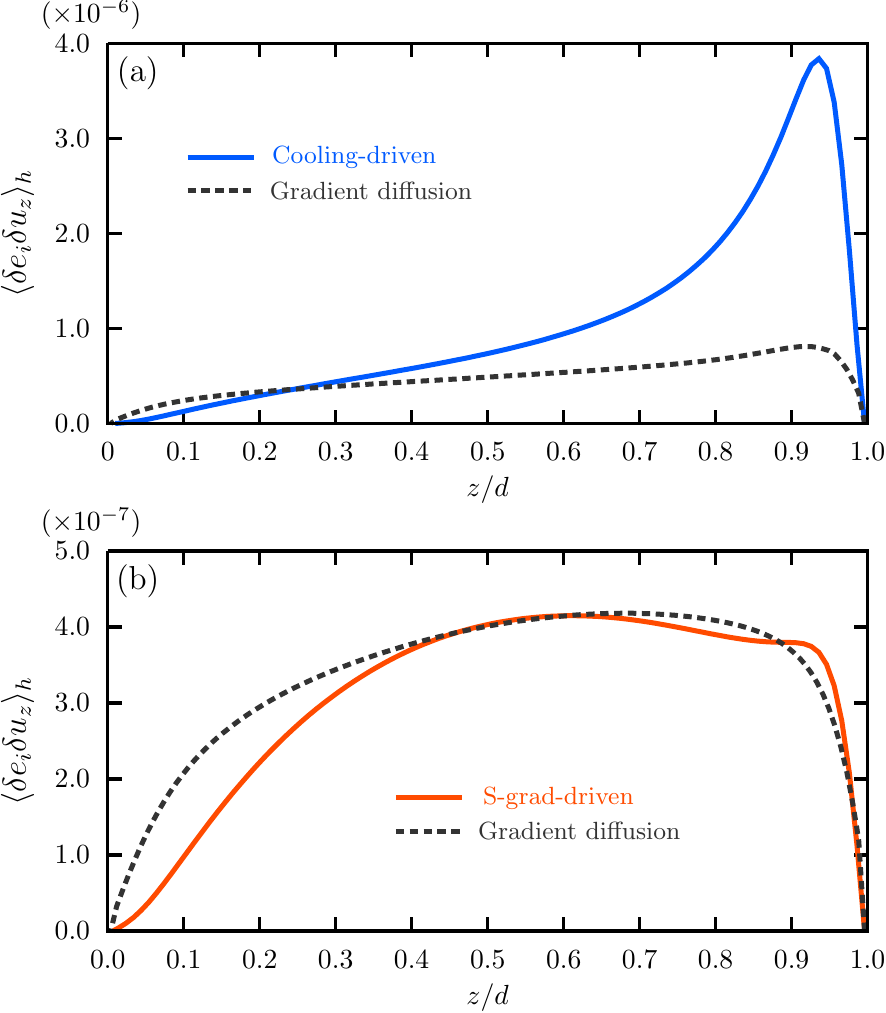}
\caption{Vertical profiles of the turbulent energy flux $\langle \delta e_i \delta u_z \rangle_{\rm h}$ from simulations (blue solid: Model~C; red dashed: Model~S) compared with predictions of the gradient-diffusion (GD) model (black dashed). (a) Cooling-driven convection (Model~C). (b) Entropy-gradient-driven convection (Model~S). While the GD approximation reproduces the simulated flux reasonably well in Model~S, it significantly underestimates the transport in the upper convection zone of Model~C, where plume-like downflows induced by surface cooling dominate the energy transport.
}
\label{fig11}
\end{figure}
\subsection{Time-Space Double Averaging (TSDA) Method}
\begin{figure}
\includegraphics[width=0.47\textwidth]{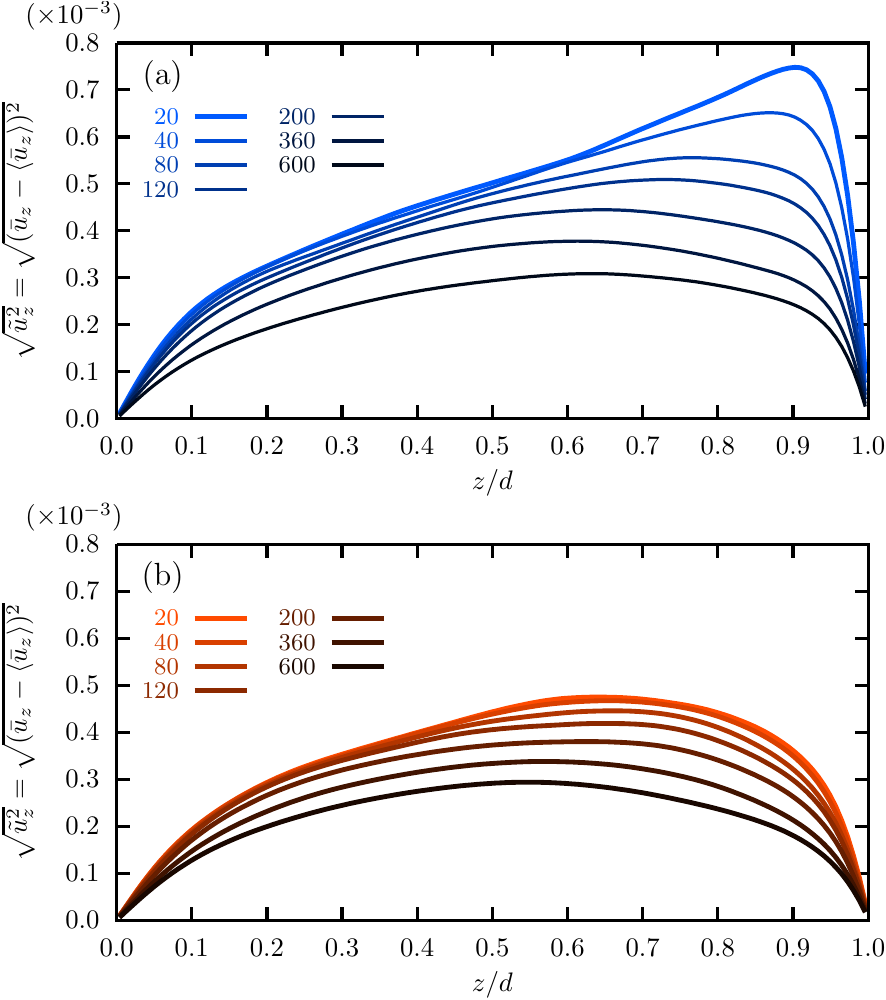}
\caption{Vertical profiles of the RMS amplitude of the coherent component of the vertical velocity, $\tilde{u}_z$, obtained using the Time–Space Double Averaging (TSDA) method for (a) Model~C and (b) Model~S. Multiple curves correspond to different averaging windows $T$ (from $T=20$ to $600$ in computational time unit): as $T$ increases, the curve color gradually shifts from blue to black in panel~(a) and from red to black in panel~(b). For reference, the convective turnover times are $\tau_{\rm cv}=16$ for Model~C and $\tau_{\rm cv}=19$ for Model~S.
}
\label{fig12}
\end{figure}
As shown in the previous subsection, the cooling-driven model (Model~C) exhibits turbulent energy fluxes that deviate strongly from GD predictions, owing to the intermittent plume-like downflows in the upper CZ. To quantify such non-equilibrium processes, we employ the Time–Space Double Averaging (TSDA) method, introduced in our Paper I \citep{yokoi+22}. See  \citet{yokoi23} for the TSDA formulation, and e.g., \citet{finnigan+08,dey14,dey+20} for the double-averaging methodology. 

In TSDA, a field quantity $f$ is first averaged in time, $\overline{f}$, and then in space, $\langle \overline{f} \rangle$. Using this procedure, $f$ is decomposed into three parts:
\begin{equation}
f = \langle \overline{f} \rangle + \tilde{f} + f^{\prime\prime} \,,
\end{equation}
where 
\begin{equation}
\tilde{f} = \overline{f} - \langle \overline{f} \rangle
\end{equation}
is the coherent fluctuation, representing spatially organized but temporally intermittent structures such as plumes, and $f^{\prime\prime}$ is the incoherent random fluctuation, i.e., the deviation from the time average. The time average $\overline{f}$ over a window of duration $T$ can be expressed in terms of a convolution with a filter function $G$ as
\begin{equation}
	\overline{f}({\bf{x}};t)
	= \int f({\bf{x}};s)\, G(t-s)\, ds .
\end{equation}
For the simplest case of a top-hat time filter, one has
\begin{equation}
	G(t-s) = \left\{ {
		\begin{array}{ll}
		1/T	& (|t-s| \le T/2), \\
		0	& \text{otherwise}.
		\end{array}
	} \right. 
\end{equation}
This decomposition enables us to separate coherent structures from random fluctuations and quantify their role in convective transport. 

Figure~12 shows the vertical profiles of the RMS amplitude of the coherent component of the vertical velocity, $\tilde{u}_z$, for (a) Model~C and (b) Model~S. The multiple curves correspond to different choices of the averaging window $T$ (from $T=20$ to $600$ in computational units): as the window length increases, the curve color gradually shifts from blue to black in panel~(a) and from red to black in panel~(b). For reference, we recall that the convective turnover time is $\tau_{\rm cv}=16$ in Model~C and $\tau_{\rm cv}=19$ in Model~S.

For Model~C, $\tilde{u}_z$ exhibits a pronounced peak in the upper CZ when short averaging times are used. This peak coincides with the strong enhancement of the turbulent energy flux (see Fig.~8), spatially localized near the surface, indicating that plume motions are the primary carriers of energy in the cooling-driven setup. 

As the averaging time increases and exceeds the typical lifetime of plumes, their coherent contribution is averaged out. The typical plume velocity is $u_z \sim 0.01$, and their travel depth is about half the depth of the convection zone ($d=1$ in our normalization). This yields a characteristic lifetime of plumes of order $(d/2)/u_z \sim 50$ in computational units. Therefore, to extract the contribution of plumes to convective transport, the averaging time should be chosen as $T \lesssim 50$. As $T$ increases further, the coherent component is smeared and the profile gradually approaches that of Model~S. 

In Model~S, by contrast, $\tilde{u}_z$ remains relatively insensitive to the averaging window, consistent with the dominance of more stationary large-scale convective eddies.

These results demonstrate that the TSDA method successfully isolates the coherent plume contribution from the total fluctuating velocity field. The correspondence between the peak of $\tilde{u}_z$ and the turbulent energy flux highlights that plume motions are the essential ingredient missing in the GD-type local closure. In this sense, $\tilde{u}_z$ serves as a key diagnostic for quantifying the plume contribution to transport and for guiding modifications to the conventional transport models. Thus, the TSDA procedure provides a practical framework for incorporating coherent structures into improved models of convective energy transport.

\subsection{Modeling of Convective Transport in the Cooling-Driven Setup: A Modified Gradient–Diffusion Approach}
In the previous subsections, we demonstrated that the conventional GD model cannot adequately describe turbulent energy transport in the cooling-driven convection (Model~C), where stochastically generated plumes dominate in the upper CZ. The TSDA analysis has revealed that the coherent component of the vertical velocity, $\tilde{u}_z$, plays a central role in mediating this transport. Motivated by these findings, and building on the theoretical framework developed in our Paper I \citep{yokoi+22}, we now construct a modified GD model that explicitly incorporates the effect of plumes through $\tilde{\bm{u}}$.

We begin by decomposing the turbulent energy flux into contributions from coherent and incoherent fluctuations. The TSDA analysis demonstrates that the coherent velocity component is particularly prominent in the upper CZ of Model~C, where plume activity is strongest. This result is fully consistent with the theoretical derivation of Paper I, which predicted that $\tilde{\bm u}$ should enter the closure expression for turbulent transport. Thus, the present numerical analysis provides direct support for the theoretical model.

To this end, we propose a modified GD closure of the form
\begin{equation}
\langle \delta e_i \delta u_z \rangle_{\rm h} \simeq -\kappa_{\rm NE} \frac{\partial \langle \overline{e}_i \rangle_{\rm h}}{\partial z} \;,
\end{equation}
where $\kappa_{\rm NE}$ is the non-equilibrium turbulent diffusivity defined as
\begin{equation}
\kappa_{\rm NE} = \kappa_{\rm E} \left[ 1 - \alpha \langle\overline{\rho} \rangle_{\rm h}^{-\beta} \left\langle (\tilde{\bm{u}}\cdot\nabla)\overline{\bm{u}^{\prime 2}}\right\rangle_{\rm h}\right] \;,
\end{equation}
with the model constants $\alpha$ and $\beta$. Here $\kappa_{\rm E}$ is the equilibrium turbulent diffusivity given in eq.~(22). In our Paper I \citep{yokoi+22}, $\beta=1/3$ was theoretically derived, but in principle $\beta$ may take other positive values depending on the detailed physical assumptions. The second term in the brackets represents the non-equilibrium modification, which arises from coherent velocity structures such as plumes. Physically, this term suggests that the inward transport of kinetic energy of incoherent fluctuations ($\propto \overline{\bm{u}^{\prime 2}}$), mediated by plumes, alters the effective coefficient of turbulent transport.

Figure~13 presents the vertical profiles of $\langle (\tilde{\bm{u}}\cdot\nabla)\overline{\bm{u}^{\prime 2}}\rangle_{\rm h}$ in the cooling-driven model (Model~C), evaluated for different averaging windows $T$. This quantity corresponds to the nonlinear correction term in eq.~(28) arising from coherent plume dynamics. In the upper CZ, it takes negative values, indicating an inward transport of incoherent kinetic energy that enhances the effective turbulent diffusivity. The strength of this correction depends on $T$: for short windows, the negative peak is strong, reflecting the contribution of individual plumes, whereas for longer windows the effect weakens as plume coherence is smeared out. These results demonstrate that the nonlinear correction is both spatially confined near the surface and strongly dependent on temporal filtering, consistent with the physical picture of plume-mediated transport.

Figure~14 compares the vertical profiles of the turbulent energy flux measured in our simulations of Model~C with those predicted by the modified GD model. In both panels, the nonlinear correction term $\langle (\tilde{\bm{u}}\cdot\nabla)\overline{\bm{u}^{\prime 2}}\rangle_{\rm h}$ in eq.~(28) is evaluated using the TSDA result with averaging window $T=40$. 

Panel~(a) corresponds to the case where both model parameters $\alpha$ and $\beta$ in eq.~(28) are treated as free parameters and optimized simultaneously by a least-squares fit to the simulation data, yielding $\alpha=3.24\times 10^6$ and $\beta=0.20$. Panel~(b) shows the case where $\beta$ is fixed at its theoretically derived value of $1/3$ (Paper I), and only $\alpha$ is adjusted to minimize the error, giving $\alpha=3.96\times 10^6$. In both cases, the modified GD model successfully reproduces both the amplitude and the depth dependence of the turbulent energy flux, in stark contrast to the conventional GD closure, which underestimates the transport by nearly an order of magnitude. 

We emphasize that, although the model formally involves three parameters, $\beta$ and $T$ are constrained by physical considerations, leaving $\alpha$ as the only adjustable scale parameter. The fact that the model reproduces both the amplitude and depth dependence of the transport under these constraints demonstrates that the improvement is not simply due to increased flexibility, but reflects the underlying physics captured by the model. This interpretation is further supported by the depth-dependent enhancement of the flux (Fig.~8) and its correlation with coherent motions (Fig.~12).

Taken together, this analysis demonstrates that the modified closure not only captures the essential role of plume dynamics identified in \S~3 and \S~4.2, but also achieves quantitative agreement with simulation data once the model parameters are appropriately specified. The consistency between the theoretical prediction $\beta=1/3$ and the simulation-based fit provides support for the applicability of the closure derived in Paper I. Furthermore, our results establish that $\tilde{\bm{u}}$ is the key diagnostic variable for modeling convective transport in the cooling-driven setup. By embedding plume dynamics into the closure, the modified GD framework successfully bridges the gap between numerical simulations and theoretical modeling, offering a physically motivated extension applicable to non-equilibrium stellar CZs.
\begin{figure}
\includegraphics[width=0.47\textwidth]{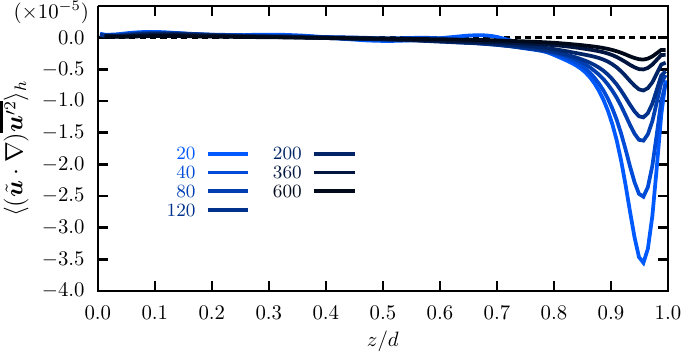}
\caption{Vertical profiles of $\langle (\tilde{\bm{u}}\cdot\nabla)\overline{\bm{u}^{\prime 2}}\rangle_{\rm h}$ in the cooling-driven model (Model~C), evaluated for different averaging windows $T$ (20, 40, 80, 120, 200, 360, and 600 in computational units). Negative values are found in the upper CZ, indicating that plume motions transport the kinetic energy of incoherent fluctuations inward. This effect enhances the effective turbulent energy transport coefficient relative to the equilibrium GD value. The magnitude of the near-surface peak depends sensitively on $T$, reflecting the role of plume lifetimes in mediating the non-equilibrium correction.}
\label{fig13}
\end{figure}

\begin{figure}
\includegraphics[width=0.47\textwidth]{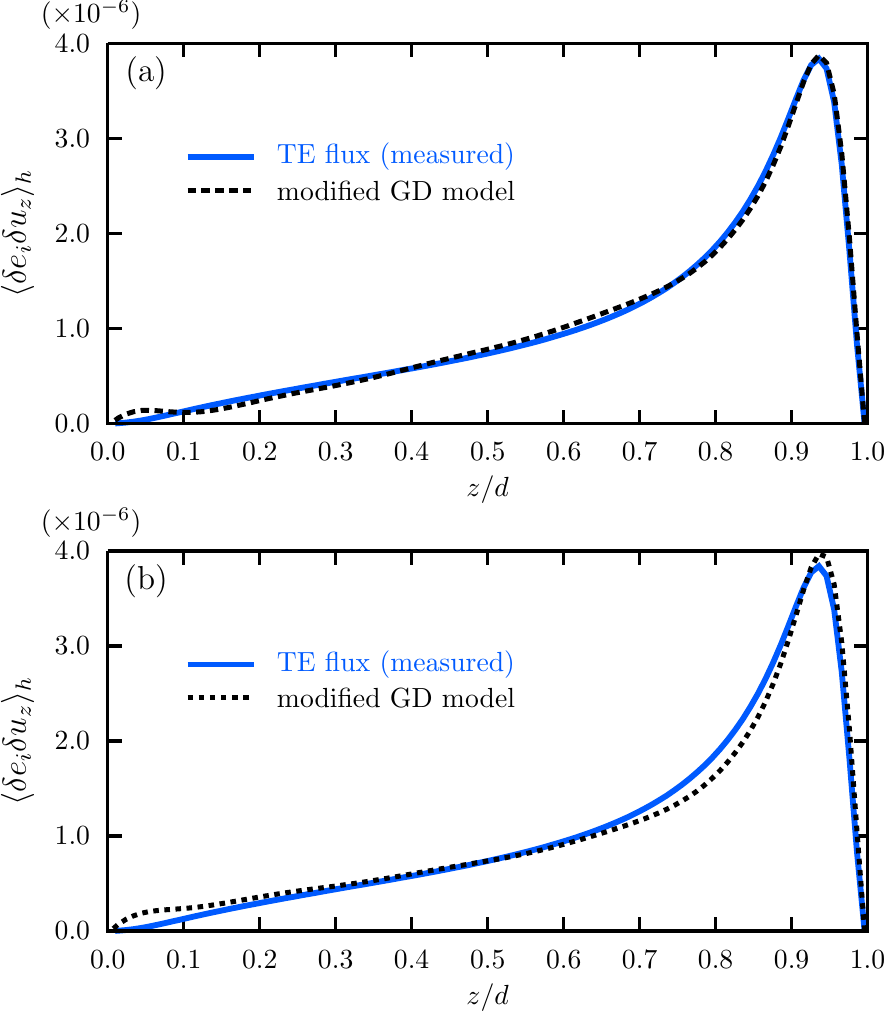}
\caption{Vertical profiles of the turbulent energy flux $\langle \delta e_i \delta u_z \rangle_{\rm h}$ in the cooling-driven model (Model~C). Blue curves show simulation results, and black dashed curves denote predictions of the modified GD model. The nonlinear correction term is evaluated using TSDA with averaging window $T=40$. (a) Fit with both $\alpha$ and $\beta$ in the non-equilibrium turbulent diffusivity defined by eq.~(28) as free parameters ($\alpha=3.24\times 10^6$, $\beta=0.20$). (b) Fit with $\beta$ fixed to its theoretical value $1/3$ and $\alpha$ optimized ($\alpha=3.96\times 10^6$). In both cases, the modified closure reproduces the measured flux far more accurately than the conventional GD model.}
\label{fig14}
\end{figure}
\section{Discussion}
The TSDA analysis together with the success of the modified closure points to a consistent physical picture: surface cooling generates coherent, intermittent plumes whose dynamics feed on—and redistribute—the energy of incoherent fluctuations. The resulting non-equilibrium transport cannot be captured by a purely local, gradient–based flux; instead, it depends on plume coherence and intermittency, encapsulated here by $\tilde{\bm{u}}$ and the correction term in eq.~(28). This picture naturally explains both the strong near-surface enhancement of the flux and its cubic scaling with downflow speed observed in Model~C.

\subsection{Limitations}
Although the modified closure reproduces key features of Model~C, two limitations remain noteworthy. 

First, the magnitude of the non-equilibrium correction, $\langle (\tilde{\bm u}\cdot\nabla)\overline{\bm u^{\prime 2}}\rangle_{\rm h}$, depends on the time-averaging window $T$ used to define coherence (Fig.~13). Physically, $T$ should be short enough to retain plume coherence (we estimated a characteristic plume lifetime $\sim 50$ in code units), yet long enough to suppress high-frequency noise. In practice this introduces a modeling choice: different $T$ values filter different fractions of the coherent contribution and thus change the inferred diffusivity. A pragmatic criterion is to select $T$ comparable to the plume turnover or travel time (estimated from $\tilde{u}_z$ and penetration depth), or to diagnose $T$ by maximizing the near-surface peak of $\tilde{u}_z$ while ensuring stationarity of the resulting profiles. 

Second, the model constant $\alpha$ (and, more generally, the exponent $\beta$) in eq.~(28) cannot yet be predicted from the fundamental equations for arbitrary conditions and must be calibrated against data (Fig.~14). Paper~I yields $\beta=1/3$ under specific assumptions and we verified that this choice is consistent with our simulations; nonetheless, the optimal $\alpha$ still depends on flow details (e.g., stratification, Rayleigh number, boundary conditions) and on the chosen $T$. This limits predictive power unless $\alpha$ can be linked to measurable statistics of $\tilde{\bm u}$ and $\overline{\bm u^{\prime 2}}$ or constrained by independent diagnostics (e.g., higher-order moments, plume frequency, or lifetimes). 

In principle, the model constants can be evaluated directly from the fundamental equations. Such an evaluation, however, requires detailed statistical information of coherent fluctuations, such as the energy $\tilde{u}^2$ and its spectral distribution. In the present work, these aspects have not been sufficiently explored. One of the main difficulties lies in the fact that the double-averaging formulation is inherently more complex than single averaging, as it involves explicit energy transfer between coherent and incoherent components of the fluctuations. Because this transfer has not yet been quantified, the model constants must for now remain adjustable parameters at the present stage.

\subsection{Astrophysical significance and modeling prospects}
Our results indicate that $\tilde{\bm{u}}$ is a key diagnostic for non-equilibrium convection, with several implications for stellar and solar modeling. First, the modified closure can be integrated into global simulations as a lightweight subgrid model, enabling large-eddy or global anelastic calculations to incorporate plume-mediated transport that is otherwise unresolved. Second, the predicted non-Gaussian statistics—including broad downflow tails, cubic flux–velocity scaling, and the near-surface localization of enhanced flux—provide concrete targets for observational tests using helioseismic inversions and high-cadence flow diagnostics.

In the broader context of the solar convection conundrum, these findings suggest a pathway toward reconciling models with observations: by explicitly accounting for coherent plumes, the modified closure may explain how the Sun sustains its luminosity despite weak large-scale convective signatures, offering a physically grounded mechanism for the near-surface intensification of turbulent energy transport \citep[e.g.,][]{hanasoge22, hotta+23, birch+24}.

\subsection{Caveats and outlook.} 
Our study neglected rotation, magnetic fields, and radiative transfer beyond a Newtonian-cooling layer, and used a Cartesian box with impenetrable, stress-free boundaries and fixed top conditions. The Prandtl and Rayleigh numbers were fixed (${\rm Pr}=1$, ${\rm Ra}=4.2\times 10^6$). Each of these choices can influence plume coherence and hence the non-equilibrium correction. Future work will survey parameter space (Ra, Pr, stratification), include rotation/MHD, test alternative boundary conditions and spherical geometry, and extend the closure to incorporate rotational alignment, magnetic tension, and radiative diffusion, thereby clarifying applicability beyond the solar CZ.

\section{Summary}
We performed 3D hydrodynamic simulations of two idealized convection setups—cooling-driven (Model~C) and entropy-gradient–driven (Model~S)—and analyzed their turbulent energy transport. Cooling-driven convection produces intermittent, plume-mediated transport that lies beyond the assumptions of local gradient–diffusion (GD) models. By isolating coherent motions with the Time–Space Double Averaging (TSDA) method and embedding their effect into a non-equilibrium diffusivity (eq.~(28)), we achieved quantitative agreement with simulations under conditions where the conventional closure fails. While the present formulation still requires an informed choice of the averaging window $T$ and a calibrated $\alpha$, it provides a physically transparent bridge between coherent plume dynamics and mean-field transport, and offers a practical pathway toward improved subgrid models for non-equilibrium stellar CZs.

Our main findings are as follows:
\begin{enumerate}
\item \textbf{Contrasting convection morphologies and thermodynamic structures:}
Model~C is dominated in the upper CZ by intermittent, plume-like downflows generated by surface cooling, whereas Model~S exhibits larger, more stationary eddies excited at depth (Figs.~3–4). These differences are also reflected in the thermodynamic stratification, with Model~C showing a stronger near-surface enhancement of superadiabaticity, whereas Model~S maintains weak superadiabaticity over a broader depth range (Fig.~5).

\item \textbf{Energy flux balance and partitioning:} 
Although the total energy flux is nearly constant with depth in both models, its partitioning differs substantially. The radiative flux provides a dominant background contribution, while the enthalpy flux is significantly enhanced in Model~C, particularly in the upper CZ, reflecting more efficient convective transport driven by surface cooling (Fig.~6).

\item \textbf{Marked non-Gaussianity in Model~C:} Probability densities of the vertical velocity develop broad downflow wings in Model~C, and the joint probability distributions of $\delta u_z$ and $\delta e_i\delta u_z$ demonstrate that frequent, high-speed downflows dominate the transport, exhibiting a nonlinear scaling $\delta e_i \delta u_z \propto |\delta u_z|^3$ in the downflow regime (Figs.~7, 9–10).

\item \textbf{Enhanced turbulent energy flux near the surface:} The vertical turbulent energy flux in Model~C exceeds that of Model~S by up to an order of magnitude in the upper CZ (Fig.~8b), despite comparable velocities at depth (Fig.~8a).

\item \textbf{Breakdown of local GD closures:} While the GD approximation reproduces Model~S fairly well, it underestimates the flux in Model~C by nearly an order of magnitude in the upper CZ (Fig.~11).

\item \textbf{Role of coherent motions and TSDA:} Using TSDA, we isolated the coherent component, $\tilde{u}_z$, and showed that it peaks where the flux is largest in Model~C. This demonstrates that plume coherence, not merely local mean gradients, mediates transport (Fig.~12).

\item \textbf{A modified non-equilibrium closure:} Building on Paper~I, we proposed a modified GD closure in which the turbulent diffusivity is corrected by a term proportional to $\langle (\tilde{\bm u}\cdot\nabla)\overline{\bm u^{\prime 2}}\rangle_{\rm h}$ (eq.~28). This term is negative in the upper CZ (Fig.~13), implying an inward transport of incoherent kinetic energy that enhances the effective diffusivity. With appropriate parameters, the modified model quantitatively reproduces both the amplitude and the depth-dependence of the flux in Model~C (Fig.~14).
\end{enumerate}
Finally, we note that the present closure still depends on the choice of $T$ and on a calibrated $\alpha$. Developing strategies to tie these parameters to observable plume statistics will be an important direction for future work.

\section*{Acknowledgements}
The authors are grateful to the anonymous referee for their insightful and constructive comments, which significantly improved the quality of the manuscript. Numerical computations were carried out on Cray XC50 at Center for Computational Astrophysics (CfCA), National Astronomical Observatory of Japan (NAOJ). This work was supported by the Research Institute of Stellar Explosive Phenomena at Fukuoka University, a Fukuoka University Research Grant (No. GR2302), and JSPS KAKENHI Grant Numbers JP17H06364, JP18H01212, JP18K03700, JP21K03612, JP21H01088, JP23K22494, JP23K25895, JP23K03400, JP24K00631, and JP25K07374. This research was also supported by MEXT as “Program for Promoting researches on the Supercomputer Fugaku” (Structure and Evolution of the Universe Unraveled by Fusion of Simulation and AI; Grant Number JPMXP1020230406) and JICFuS. The National Institutes of Natural Sciences (NINS) program for cross-disciplinary study (Grant Numbers 01321802 and 01311904) on Turbulence, Transport, and Heating Dynamics in Laboratory and Solar/Astrophysical Plasmas: “SoLaBo-X” supports this work.
\section*{Data Availability}
The data underlying this article will be shared on reasonable request to the corresponding author. 



\bibliographystyle{mnras}
\bibliography{example} 



\appendix

\section{THERMAL RELAXATION}
\begin{figure}
\includegraphics[width=0.47\textwidth]{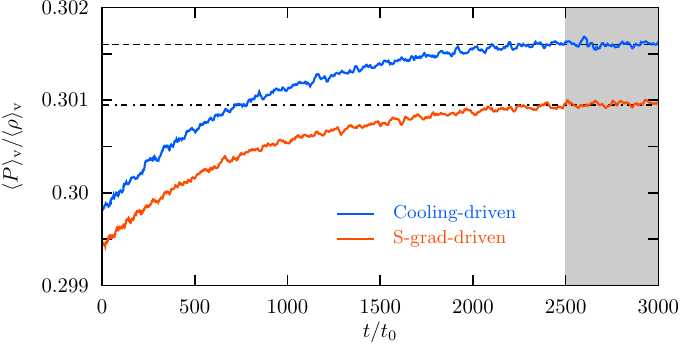}
\caption{Temporal evolution of \(\langle P\rangle_V/\langle\rho\rangle_V\) for Models C (blue) and S (red). The shaded region indicates the interval \(2500\le t/t_0\le3000\) used for the statistical analysis in our study. The horizontal dashed and dash-dotted lines indicate the time-averaged values over this interval for both models.}
\label{figA1}
\end{figure}
To assess the thermal relaxation of the simulations, Figure A1 shows the temporal evolution of \(\langle P\rangle_V/\langle\rho\rangle_V\) for Models C and S, where \(\langle\cdot\rangle_V\) denotes a volume average. For the ideal-gas equation of state adopted in our simulations, this quantity provides a measure of the mean thermal state of the system. In both models, \(\langle P\rangle_V/\langle\rho\rangle_V\) evolves gradually during the early phase and approaches a nearly stationary value at late times. During the interval \(2500\le t/t_0\le3000\), used for the statistical analysis in this study, the quantity fluctuates around an approximately constant mean value, indicating that both simulations have reached a quasi-steady thermal state.
\section{Flow asymmetry and contributions to turbulent transport}
\begin{figure}
\includegraphics[width=0.47\textwidth]{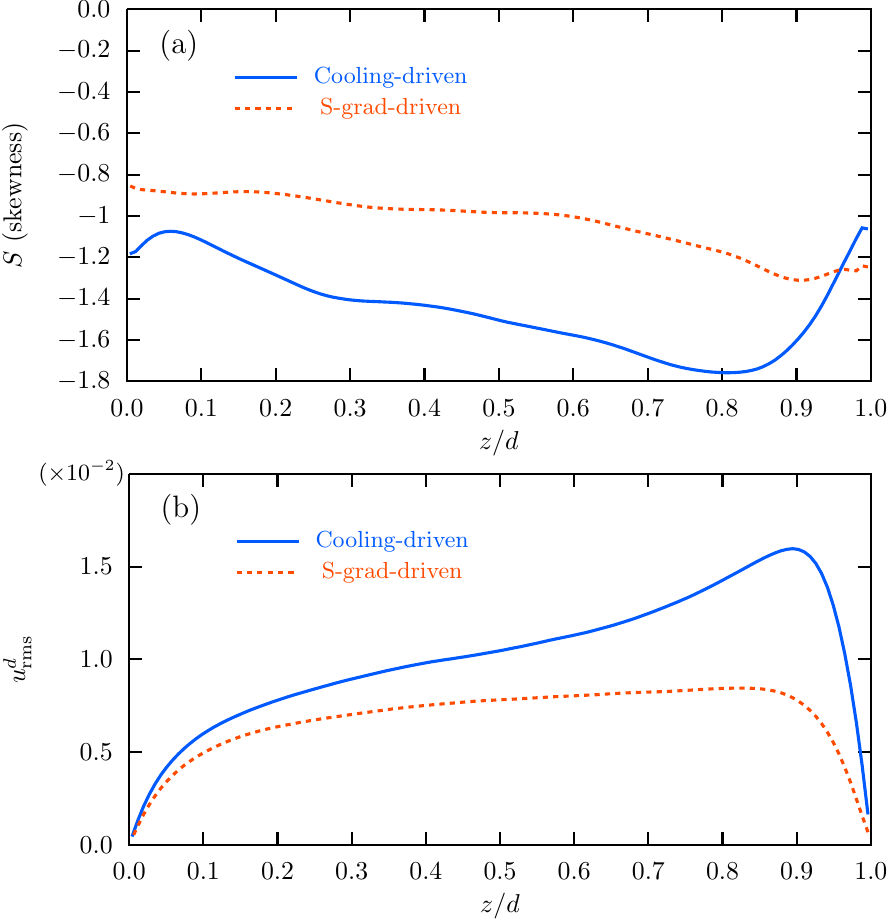}
\caption{(a) Vertical profiles of the skewness of the vertical velocity for Model~C (blue solid) and Model~S (red dashed). Negative skewness indicates the dominance of downflows. (b) Vertical profile of the root-mean-square velocity in the downflow region (downflow $u_{\rm rms}: u^d_{{\rm rms}}$) for Model~C (blue solid) and Model~S (red dashed).}
\label{figB1}
\end{figure}
\begin{figure}
\includegraphics[width=0.47\textwidth]{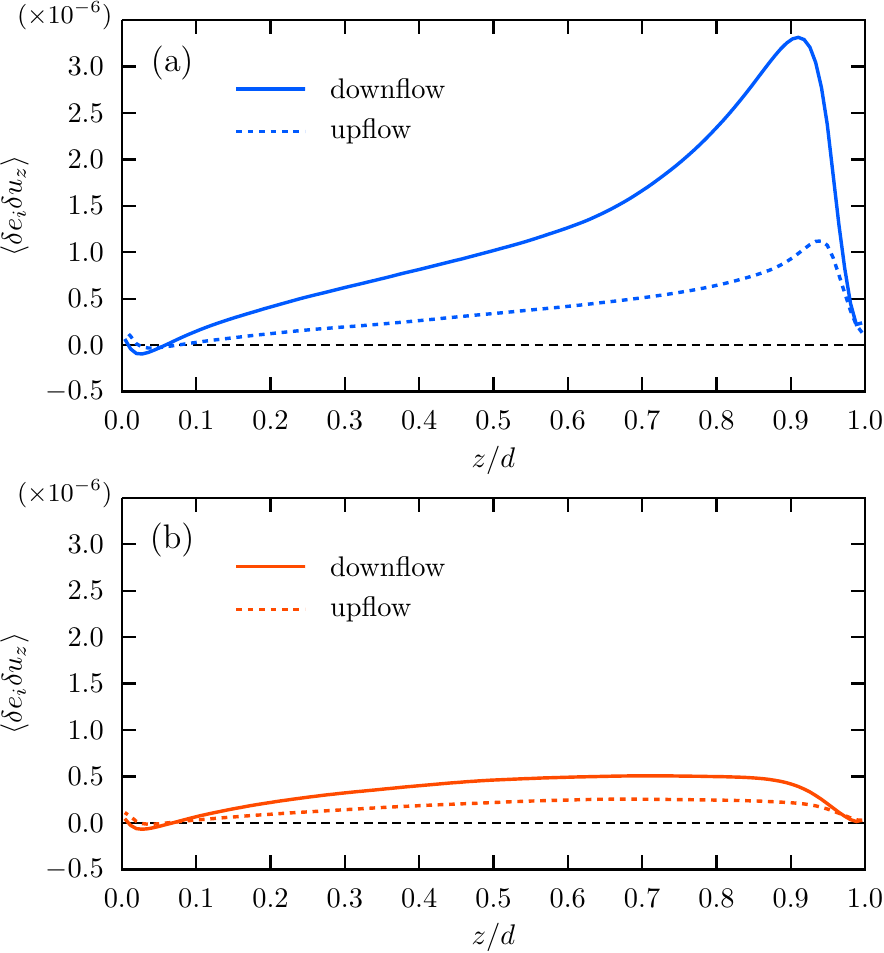}
\caption{Vertical profiles of the turbulent energy flux $f_{\rm turb} (z) = \langle\langle \delta e_i \, \delta u_z \rangle_{\rm h}\rangle$, decomposed into contributions from downflows ($f_{\rm turb}^{d}$: blue solid) and upflows ($f_{\rm turb}^{u}$: red dashed), for (a) Model~C and (b) Model~S.}
\label{figB2}
\end{figure}
To quantify the flow asymmetry and its impact on energy transport, we introduce the following diagnostics. Again, $\langle \cdot \rangle_{\rm h}$ and $\langle \cdot \rangle$ denote horizontal and temporal averages, respectively, and fluctuations are defined with respect to the horizontal mean as $\delta f = f - \langle f \rangle_{\rm h}$.

The skewness of the vertical velocity is defined as
\begin{equation}
S(z) = \frac{\left\langle\left\langle (\delta u_z)^3 \right\rangle_{\rm h}\right\rangle}
{\left\langle\left\langle (\delta u_z)^2 \right\rangle_{\rm h}\right\rangle^{3/2}} \;.
\end{equation}
This quantity measures the asymmetry of the velocity distribution, with negative values indicating the dominance of strong, narrow downflows over broader, weaker upflows.

We decompose the flow into upflow and downflow regions based on the sign of the vertical velocity. The downflow region is defined by $u_z < 0$ and the upflow region by $u_z > 0$. Then, the root-mean-square velocity in the downflow region (downflow $u_{\rm rms}: u^d_{{\rm rms}}$) is defined as
\begin{equation}
u^d_{{\rm rms}}(z) = \left\langle \left\langle (\delta u_z)^2 \,\Theta(-u_z) \right\rangle_{\rm h} \right\rangle^{1/2},
\end{equation}
and that in the upflow region (upflow $u_{\rm rms}: u^u_{{\rm rms}}$) is 
\begin{equation}
u^u_{{\rm rms}}(z) = \left\langle \left\langle (\delta u_z)^2 \,\Theta(u_z) \right\rangle_{\rm h} \right\rangle^{1/2},
\end{equation}
where $\Theta(x)$ is the Heaviside step function, defined as $\Theta(x)=1$ for $x>0$ and $\Theta(x)=0$ otherwise. 

The turbulent energy flux is defined as
\begin{equation}
f_{\rm turb}(z) = \left\langle\left\langle \delta e_i \, \delta u_z \right\rangle_{\rm h}\right\rangle,
\end{equation}
and decomposed into contributions from downflows and upflows as
\begin{equation}
f_{\rm turb}^{d}(z) = \left\langle \left\langle \delta e_i \, \delta u_z \, \Theta(-u_z) \right\rangle_{\rm h}
\right\rangle,
\end{equation}
and 
\begin{equation}
f_{\rm turb}^{u}(z) = \left\langle \left\langle \delta e_i \, \delta u_z \, \Theta(u_z) \right\rangle_{\rm h} \right\rangle,
\end{equation}
such that $f_{\rm turb} = f_{\rm turb}^{d} + f_{\rm turb}^{u}$.

This decomposition allows us to directly quantify the relative contributions of upflows and downflows to the turbulent transport.

To further clarify the origin of the enhanced enthalpy flux in Model~C, we examine the statistical properties 
of the convective flows and the contributions from upflows and downflows.

Figure~B1(a) shows the vertical profile of the skewness of the vertical velocity. In both models, the skewness is negative throughout most of the convection zone, indicating a dominance of narrow, fast downflows and broader, slower upflows. This asymmetry is significantly stronger in Model~C, reflecting the effect of localized surface cooling. 

Figure~B1(b) shows the root-mean-square velocity in the downflow region for Model~C (blue solid) and Model~S (red dashed). The cooling-driven model exhibits systematically larger velocities, particularly in the upper layers, consistent with stronger convective driving.

To directly assess the contribution to heat transport, Fig. B2 shows the turbulent energy flux \(f_{\rm turb}\), decomposed into contributions from upflows and downflows. The results demonstrate that the transport is dominated by downflows in both models, but the dominance is much stronger in Model~C. The enhanced contribution from downflows explains the larger enthalpy flux observed in Fig.~6.

These results indicate that the efficiency of convective heat transport is controlled not only by the overall flow amplitude but also by the asymmetry of the flow and the correlation between thermodynamic fluctuations and vertical velocity. The cooling-driven setup promotes strong, coherent downflows that enhance this correlation, leading to more efficient convective energy transport.

\bsp	
\label{lastpage}
\end{document}